\documentclass[fleqn,usenatbib]{mnras}

\usepackage[T1]{fontenc}
\usepackage{ae,aecompl}

\usepackage{newtxtext,newtxmath}

\usepackage{graphicx}
\usepackage{amsmath}

\usepackage{aas_macros}
\usepackage{multirow}
\usepackage{orcidlink}
\usepackage[normalem]{ulem}

\DeclareRobustCommand{\VAN}[3]{#2}
\let\VANthebibliography\thebibliography
\def\thebibliography{\DeclareRobustCommand{\VAN}[3]{##3}\VANthebibliography}

\title[The multiwavelength structure of PSBs]{The multiwavelength structure of post-starburst galaxies at $0.5 < z < 3$ with {\emph{JWST}} PRIMER: compact morphologies and residual disturbances}
\author[D.~T.~Maltby et al.]
{David~T.~Maltby\textsuperscript{\,\orcidlink{https://orcid.org/0000-0002-8163-080X}},$^{1}$\thanks{E-mail: david.maltby@nottingham.ac.uk}
Omar~Almaini\textsuperscript{\,\orcidlink{0000-0001-9328-3991}},$^{1}$
Vivienne~Wild\textsuperscript{\,\orcidlink{0000-0002-8956-7024}},$^{2}$
Elizabeth~Taylor\textsuperscript{\,\orcidlink{0000-0001-8728-2984}},$^{3}$
Kate~Rowlands\textsuperscript{\,\orcidlink{0000-0001-7883-8434}},$^{4,5}$
\newauthor Thomas~de~Lisle,$^{1}$
Guillaume~Hewitt\textsuperscript{\,\orcidlink{0009-0006-7827-007X}},$^{1}$
Maya Skarbinski\textsuperscript{\,\orcidlink{0009-0004-0844-0657}},$^{4}$
James~S.~Dunlop\textsuperscript{\,\orcidlink{0000-0002-1404-5950}},$^{3}$
Adam~C.~Carnall\textsuperscript{\,\orcidlink{0000-0002-1482-5818}},$^{3}$
\newauthor Anton~M.~Koekemoer\textsuperscript{\,\orcidlink{0000-0002-6610-2048}},$^{6}$
and Derek~J.~McLeod\textsuperscript{\,\orcidlink{0000-0003-4368-3326}}$^{3}$\\
$^{1}$School of Physics and Astronomy, University of Nottingham, University Park, Nottingham NG7 2RD, UK\\
$^{2}$School of Physics and Astronomy, University of St Andrews, North Haugh, St Andrews KY16 9SS, UK\\
$^{3}$Institute for Astronomy, University of Edinburgh, Royal Observatory, Edinburgh EH9 3HJ, UK\\
$^{4}$William~H.~Miller III Department of Physics and Astronomy, Johns Hopkins University, Baltimore, MD 21218, USA\\
$^{5}$AURA for ESA, Space Telescope Science Institute, 3700 San Martin Drive, Baltimore, MD 21218, USA\\
$^{6}$Space Telescope Science Institute, 3700 San Martin Drive, Baltimore, MD 21218, USA}

\date{Accepted 2026 May 26. Received 2026 April 29; in original form 2026 January 19}

\pubyear{000}

\begin{document}

\label{firstpage}

\pagerange{\pageref{firstpage}--\pageref{lastpage}}

\maketitle


\begin{abstract}
We investigate the multi-wavelength structure of recently-quenched post-starburst (PSB) galaxies at $0.5 < z < 3$, using photometrically selected samples from the Ultra Deep Survey (UDS).  Leveraging deep eight-band \emph{JWST}/NIRCam imaging from the PRIMER programme, we analyze $\sim120$ PSBs across the rest-frame optical–to-near-infrared, and compare with a reference sample of $\sim3000$ passive and star-forming galaxies.  Structural parameters (effective radius $R_{\rm e}$ and S\'ersic index~$n$) are derived independently in each waveband, and reveal that PSBs exhibit minimal structural variation with wavelength, indicating negligible stellar population age gradients or internal dust obscuration.  We confirm that PSBs follow the established redshift–mass trends: at $z > 1$, massive PSBs ($M_* > 10^{10}\,{\rm M}_\odot$) are compact spheroids resembling massive passive galaxies, albeit significantly more compact, whereas at $0.5 < z < 1$, PSBs are typically low-mass ($M_* < 10^{10}\,{\rm M}_\odot$) compact, disc-dominated systems akin to low-mass passive discs.  Furthermore, for the first time, we systematically quantify disturbance indicators (residual flux fraction $\textit{RFF}$, asymmetry, residual asymmetry) across a large PSB sample.  At all masses, PSBs exhibit low $\textit{RFF}$ and asymmetry values comparable to passive systems and consistent with smooth, largely undisturbed morphologies.  However, at $z > 1$, massive PSBs ($M_* > 10^{10.25}\,{\rm M}_\odot$) show enhanced residual asymmetry relative to the passive population, indicating a previously unrecognized level of structural disturbance masked beneath a smooth stellar distribution.  These results suggest that, while structural transformation is largely complete by the PSB phase, residual disturbances persist at high redshift, supporting a scenario in which rapid quenching proceeds via two distinct pathways: highly disruptive events (e.g. major mergers) at high $z$ and high mass, and comparatively gentle processes at later times.
\end{abstract}

\begin{keywords}
galaxies: evolution --- galaxies: fundamental parameters --- galaxies: high-redshift --- galaxies: structure
\vspace{-0.1cm}
\end{keywords}

\section{Introduction}

\label{Introduction}

In the local Universe, the galaxy population exhibits a pronounced bimodality in optical colour, star-formation activity, and morphology \citep[e.g.][]{Strateva_etal:2001, Schawinski_etal:2014}.  Massive galaxies are predominantly red, quiescent systems with early-type morphologies, whereas lower-mass galaxies are typically blue, actively star-forming, and display late-type structures. These two groups correspond to the \emph{red sequence} and \emph{blue cloud}, respectively.  Since $ z > 2$, this bimodality has evolved substantially, with a rapid build-up of stellar mass on the red sequence driven by the quenching of blue cloud galaxies \citep[e.g.][]{Bell_etal:2004, Cirasuolo_etal:2007, Faber_etal:2007, Brammer_etal:2011, Ilbert_etal:2013, Muzzin_etal:2013}.  Despite significant progress, however, the physical mechanisms responsible for this quenching remain poorly understood.

A wide range of mechanisms have been proposed to explain the quenching of star formation at high redshift.  These processes generally act by removing cold gas (the raw material for star formation) and/or stabilizing it against collapse through the injection of heat and turbulence.  They include gas-stripping processes \citep{Gunn&Gott:1972}, shock heating of infalling gas by a massive hot halo \citep{Dekel&Birnboim:2006}, morphological quenching \citep{Martig_etal:2009}, and the gradual depletion of the gas reservoir through strangulation \citep{Larson_etal:1980}.  Quenching may also be driven by feedback processes, in which AGN- or starburst-powered outflows expel the cold gas required for sustained star formation \citep[e.g.][]{Silk&Rees:1998, Hopkins_etal:2005, Diamond-Stanic_etal:2012}.  In addition, radio-mode AGN activity may suppress further gas accretion, helping to maintain quiescence on longer time-scales \citep{Best_etal:2005, Best_etal:2006}.  Collectively, these processes are often categorized into \emph{rapid} and \emph{slow} quenching pathways and interpreted within the framework of \emph{mass quenching}, governed by internal processes linked to stellar mass, and \emph{environmental quenching}, driven by external influences.  Observational evidence suggests that slow quenching dominates in the local Universe \citep[e.g.][]{Peng_etal:2015}, whereas rapid quenching becomes increasingly important at $z > 1$ \citep[][]{Barro_etal:2013, Wild_etal:2016, Carnall_etal:2018, Belli_etal:2019}.

The cessation of star formation in massive galaxies is also accompanied by significant changes in their structural properties.  At $z\sim2$, massive galaxies are predominantly disc-dominated systems, whereas their present-day counterparts are largely spheroidal \citep{vanderWel_etal:2011, Buitrago_etal:2013}.  This morphological transformation appears to occur primarily at $z > 1$ for galaxies with $M_* > 10^{10.5}\,\rm{M_{\odot}}$ \citep{Mortlock_etal:2013}.  However, it remains unclear whether this structural transformation coincides with, precedes, or follows the quenching of star formation itself.

Massive galaxies at early epochs are also observed to be significantly more compact than systems of comparable stellar mass in the local Universe \citep[e.g.][]{Trujillo_etal:2006}.  This compactness implies substantial size evolution, often attributed to growth through minor mergers \citep{Naab_etal:2009}, although alternative interpretations have been proposed \citep[e.g. progenitor bias;][]{Carollo_etal:2013}.  Several mechanisms have been suggested for the formation of these compact high-redshift systems (i.e.\ \emph{red nuggets}).  For example, they may arise through a highly dissipative process, such as gas-rich mergers \citep{Hopkins_etaL:2009, Wellons_etal:2015} or protogalactic collapse \citep{Dekel_etal:2009, Zolotov_etal:2015}, followed by rapid quenching via AGN or starburst-powered outflows \citep[e.g.][]{Hopkins_etal:2005}.  Alternatively, they could have formed at very early times when the Universe itself was much denser \citep{Wellons_etal:2015}.

To probe the mechanisms driving quenching and structural evolution at high redshift, it is instructive to study galaxies that have been quenched recently and are therefore observed in transition.  Post-starburst (PSB) galaxies constitute a rare population that underwent a major burst of star formation, rapidly truncated within the last few hundred Myr.  They are spectroscopically identified by strong Balmer absorption features, indicative of a dominant A-type stellar population, and by the absence of strong nebular emission lines \citep{Dressler&Gunn:1983, Wild_etal:2009}.  Owing to their intrinsically short-lived nature, only a limited number of PSBs have been spectroscopically confirmed at $z > 1$.  Significant progress has, however, been made through photometric techniques, enabling the identification of much larger samples of high-redshift PSBs.  For example, \cite{Whitaker_etal:2012a} used medium-band near-infrared photometry to identify `young red-sequence' galaxies from rest-frame $\mathit{UVJ}$ colour--colour diagrams, while \cite{Wild_etal:2014} developed a classification scheme based on Principal Component Analysis (PCA) of broad-band galaxy SEDs. This PCA technique has been shown to be particularly effective and robustly verified using deep optical spectroscopy \citep{Maltby_etal:2016}.  Applied to photometric data from the Ultra Deep Survey, it has led to the identification of $\sim2000$ PSBs across $0.5 < z < 3$ \citep{Wilkinson_etal:2021}.

Using PSBs as a window onto quenching, observational evidence now indicates that PSBs comprise at least two distinct sub-populations, each linked to a different quenching pathway.  This conclusion is supported by stellar mass functions (\citealt{Wild_etal:2016}; de Lisle et al., in preparation), structural analyses \citep{Maltby_etal:2018}, environmental trends \citep{Taylor_etal:2023}, and measurements of large-scale structure \citep{Wilkinson_etal:2021}.  At $0.5 < z < 1$, PSBs are typically low-mass ($M_* < 10^{10}\,\rm{M_{\odot}}$), compact, disc-dominated systems that are preferentially found in over-dense environments \citep{Maltby_etal:2018, Socolovsky_etal:2018, Socolovsky_etal:2019, Taylor_etal:2023}.  This suggests rapid quenching via relatively gentle, environmentally driven processes that preserve disc structure, such as minor mergers or ram-pressure stripping.  In contrast, at $z > 1$, massive PSBs ($M_* > 10^{10}\,\rm{M_\odot}$) are extremely compact and spheroidal \citep[e.g.][]{Whitaker_etal:2012a, Almaini_etal:2017, Maltby_etal:2018, Zhang_etal:2024, Clausen_etal:2024, Clausen_etal:2025}, consistent with rapid, highly dissipative quenching via major mergers or protogalactic collapse.  Massive PSBs at $z > 1$ also show evidence for high-velocity outflows \citep{Maltby_etal:2019, Taylor_etal:2024}, consistent with rapid, internally driven quenching, potentially linked to AGN feedback.  Although, strong AGN signatures are rarely observed, suggesting a short AGN duty cycle or significant obscuration (\citealt{French_etal:2023}; \citealt{Almaini_etal:2025}; \citealt{Krishna_etal:2025}; Hewitt et al., in preparation).  Together, these results support a dual-channel scenario for rapid quenching, with distinct pathways operating at different epochs and/or mass regimes.  Similar conclusions are also reached by recent studies using the older passive population at $z > 1$, which report a flattening of the size--mass relation at low stellar masses and evidence for a mass-dependent bimodality \citep[e.g.][]{Nedkova_etal:2021, Cutler_etal:2024, Hamadouche_etal:2025, Miller_etal:2026}.  In particular, \cite{Cutler_etal:2024} and \cite{Hamadouche_etal:2025} also report a corresponding morphological bimodality for quiescent galaxies, consistent with different formation and quenching pathways for low and high-mass systems.

Despite this coherent picture, it is important to note that much of the structural evidence relies on measurements over a narrow wavelength range and may be affected by dust obscuration or stellar population gradients.  For instance, \cite{Suess_etal:2020} argued that the apparent compactness of PSBs at $z > 1$ could be driven by colour gradients, since younger quiescent galaxies exhibit flatter gradients than older systems \citep[see also][]{Suess_etal:2022}.  This highlights the need for multiwavelength analyses, particularly in the rest-frame near-infrared, to robustly trace the underlying stellar mass distribution.  Furthermore, in the nearby Universe, rapidly quenched galaxies are strongly associated with post-merger signatures \citep[e.g.][]{Pawlik_etal:2016, Pawlik_etal:2018, Wilkinson_etal:2022, Ellison_etal:2022, Ellison_etal:2024, Gordon_etal:2026}, and simulations suggest such features can remain detectable for up to $\sim1\,\rm{Gyr}$ at high redshift \citep{ Whitney_etal:2021}.  However, the prevalence of post-merger signatures among rapidly-quenched galaxies at higher redshift remains, as of yet, unexplored.

In this paper, we explore the multiwavelength structural properties of PSBs over the redshift range $0.5 < z < 3$, using the sample of \cite{Wilkinson_etal:2021} together with eight-band \emph{JWST}/NIRCam imaging from Public Release IMaging for Extragalactic Research (PRIMER; Dunlop et al., in preparation).  Our primary aim is to assess the impact of stellar age gradients and dust obscuration on previous measurements of PSB structure, which were derived over a limited wavelength range \citep[$\lambda_{\rm obs} \leq 2.2\,\mu\rm{m}$; e.g.][]{Whitaker_etal:2012a, Almaini_etal:2017, Maltby_etal:2018, Zhang_etal:2024}.  We build on these studies by (i) extending PSB structural analyses significantly further into the near-infrared ($\lambda_{\rm obs} > 2.2\,\mu\rm{m}$), and (ii) quantifying structural disturbance in PSBs across $0.5 < z < 3$, to assess the prevalence of residual merger signatures within the PSB population.

The structure of this paper is as follows.  In Section~\ref{Data and sample selection}, we provide a brief description of the data relevant to this work and outline our sample selection.  In Section~\ref{Measuring structural parameters}, we describe the measurement of galaxy structural parameters in the UDS field using the PRIMER near-infrared imaging.  In Section~\ref{The structure of post-starburst galaxies}, we present an analysis of the multiwavelength structure of star-forming, passive, and PSB galaxies across the redshift range $0.5 < z < 3$, complemented by an exploration of their disturbance metrics in Section~\ref{Signatures of disturbance}.  We include a discussion of our results and draw our conclusions in Section~\ref{Conclusions}.  Throughout this paper, we use AB magnitudes and adopt a cosmology of $H_{0} = 70\,\mathrm{km\,s^{-1}\,Mpc^{-1}}$, $\Omega_{\Lambda} = 0.7$, and $\Omega_{\mathrm{m}} = 0.3$.

\section{Data and sample selection}

\label{Data and sample selection}

This work is based on galaxy populations identified from the multiwavelength photometry of the final UDS data release (DR11).  However, the structural parameters used throughout are derived from the recent \emph{JWST}/NIRCam imaging of the UDS field, obtained as part of the PRIMER survey (Dunlop et al.\ in preparation).  We use the galaxy sample from \cite{Wilkinson_etal:2021} to enable direct comparison with previous UDS PSB studies, with the deeper \emph{JWST} imaging being used for sample selection in forthcoming work (de Lisle et al., in preparation).  A brief overview of these data is provided in the following sections.

\subsection{The Ultra Deep Survey (UDS)}

\label{The UDS}

The UDS is the deepest component of the UKIRT Infrared Deep Sky Survey \citep[UKIDSS;][]{Lawrence_etal:2007}, providing exceptionally deep $JHK$ photometry over an area of $0.77\,\mathrm{deg}^2$.  The final data release (DR11; June 2016) achieved $5\sigma$ limiting depths of $J = 24.9$, $H = 24.2$, and $K = 25.3$ (AB; 2-arcsec apertures), making it the deepest contiguous $K$-band survey over such an area to date.

The UDS benefits from extensive complementary multiwavlength coverage, including deep optical $BVRi'z'$ photometry from the Subaru--{\emph{XMM-Newton}} Deep Survey \citep[SXDS; ][]{Furusawa_etal:2008}, and mid-infrared observations at $3.6$ and $4.5\,\mu\mathrm{m}$ from a combination of the  \emph{Spitzer} UDS Legacy Program (SpUDS; PI:~Dunlop) and the \emph{Spitzer} Extended Deep Survey \citep[SEDS;][]{Ashby_etal:2013}.  Additional data include deep $u'$-band imaging from MegaCam on the Canada--France--Hawaii Telescope, and $Y$-band data from the VISTA VIDEO survey \citep{Jarvis_etal:2013}.  The full multiwavelength coverage, after masking bright stars and artifacts, spans approximately $0.62\,\mathrm{deg}^2$ of the UDS field \citep[see][for further details]{Almaini_etal:2017, Almaini_etal:2025}.

For the UDS photometric catalogues, source detection and multiwavelength photometry were performed using {\sc SExtractor} \citep{Bertin&Arnouts:1996} in dual-image mode, with the $K$-band as the detection image.  Photometry was extracted in 2-arcsec apertures and subsequently corrected to total fluxes. Stars and image artifacts were identified and excluded, resulting in a final galaxy catalogue comprising approximately $205\,000$ galaxies.

Photometric redshifts are determined using the {\sc eazy} code \citep{Brammer_etal:2008}, which fits the 12-band photometry with a grid of galaxy templates that span a wide range of stellar population ages, including dust-reddened SEDs.  The default template set of 12 Flexible Stellar Population Synthesis (FSPS) models \citep{Conroy&Gunn:2010} is supplemented with three additional simple stellar population (SSP) templates from \citet{Bruzual&Charlot:2003} to cover recent star formation activity.  These additional SSP templates have ages of 20, 50, and 150 Myr, assume a \citet{Chabrier:2003} initial mass function (IMF), and adopt a sub-solar metallicity of 0.2~$Z_\odot$.  The photometric redshifts were calibrated against a sample of approximately $8000$ sources with secure spectroscopic redshifts, with the comparison yielding a normalised median absolute deviation of $\sigma_{\rm NMAD} = 0.019$ in $\Delta z / (1 + z)$ and an outlier fraction [$\Delta z / (1 + z) > 0.15$] of $\sim3$ per cent.  For further details, see \citet{Taylor_etal:2023}.

\subsection{The PRIMER Survey}

\label{PRIMER}

In this work, we utilize the deep near-infrared imaging from a key \emph{JWST} survey: Public Release IMaging for Extragalactic Research (PRIMER; Dunlop et al., in preparation).  PRIMER is a Cycle 1 Treasury Programme that provides deep \emph{JWST} NIRCam and MIRI imaging of two \emph{HST} CANDELS legacy fields --- COSMOS and UDS.  The NIRCam imaging comprises eight filters ($F090W$, $F115W$, $F150W$, $F200W$, $F277W$, $F356W$, $F410M$ and $F444W$) which span the wavelength range $0.9$--$4.4\,\mu\mathrm{m}$ and cover a total area of $\sim380\,\mathrm{arcmin}^2$ ($\sim144\,\mathrm{arcmin}^2$ for PRIMER-COSMOS and $\sim234\,\mathrm{arcmin}^2$ for PRIMER-UDS).  Complementary MIRI observations are available in two filters ($F770W$ and $F1800W$) which probe wavelengths of $7.7\,\mu\mathrm{m}$ and $18\,\mu\mathrm{m}$, respectively, and cover a total area of $\sim240\,\mathrm{arcmin}^2$ ($\sim112\,\mathrm{arcmin}^2$ for PRIMER-COSMOS and $\sim125\,\mathrm{arcmin}^2$ for PRIMER-UDS).  As \emph{HST} legacy fields, both PRIMER regions also benefit from the deep optical/ACS imaging of the CANDELS survey \citep{Grogin_etal:2011, Koekemoer_etal:2011}, providing observations in three additional filters ($F435W$, $F606W$ and $F814W$) which cover the wavelength  range $0.4$--$0.8\,\mu\mathrm{m}$.  In this study, we make use of the PRIMER-UDS NIRCam observations for our morphological analyses, which cover $\sim8$ per cent of the UDS field.

\subsection{Sample selection}

\label{Sample selection}

In this work, we use a large sample of galaxies from UDS DR11 in the redshift range $0.5 < z < 3$, classified by \cite{Wilkinson_etal:2021}.  Their classification is based on the photometric method developed by \cite{Wild_etal:2014}, which employs a principal component analysis (PCA) of galaxy SEDs.  This method describes a wide variety of SED shapes through a linear combination of a small number of principal components.  These components are derived from a large library of `stochastic burst' model SEDs generated using \cite{Bruzual&Charlot:2003} stellar population synthesis models with stochastic star formation histories.  The approach yields a mean SED, $m_{\lambda}$, and a series of $p$ eigenspectra $e_{i\lambda}$ (i.e.\ principal components), such that any normalised SED, $f_{\lambda}/n$, can be approximately reconstructed as
\begin{equation}
\frac{f_{\lambda}}{n} = m_{\lambda} + \sum^{p}_{i=1} a_ie_{i\lambda},
\end{equation}
where the amplitudes $a_i$ are termed supercolours (SC).  It turns out that only the first three SCs are required to provide a compact representation of a wide variety of SED shapes, accounting for $>99.9$ per cent of the variance in the models of \cite{Wild_etal:2014}.

These SCs correlate with key physical properties including mean stellar age, dust content, metallicity, and the burst mass fraction formed within the last Gyr.  These correlations enable a robust classification of galaxies into star-forming, passive, dusty star-forming and PSB populations.  PSBs are identified as galaxies that have undergone recent and rapid quenching, having formed more than $10$ per cent of their stellar mass within the last Gyr.  However, this selection does not necessarily imply a short, intense burst of star formation; some PSBs may instead have experienced a more extended period of elevated star formation that was nonetheless rapidly quenched \citep[see][for further details]{Wild_etal:2016, Wild_etal:2020}.  

For observed galaxies, SCs are determined by projecting their SEDs onto the PCA eigenspectra.  A key advantage of this approach is that it does not rely directly on model fitting, and it allows for the classification of SEDs that lie outside the range covered by the models.  In the UDS field, this analysis is applied to all galaxies with $K_{\rm AB} < 24.5$ and $0.5 < z < 3$ \citep{Wilkinson_etal:2021}, resulting in a sample of $81\,102$ star-forming, $5102$ passive, and $1943$ PSB galaxies.  Note that throughout this work, the dusty star-forming class is incorporated into the broader star-forming population.

In \cite{Wilkinson_etal:2021}, stellar masses are estimated using a Bayesian analysis, following the method outlined in \cite{Wild_etal:2016}.  Briefly, a large library of \cite{Bruzual&Charlot:2003} population synthesis models is constructed, incorporating a wide range of star-formation histories.  These models are then fitted to the galaxy SCs, yielding probability distributions for various physical parameters, including stellar mass.  The masses assume a \cite{Chabrier:2003} initial mass function (IMF), and have typical random uncertainties of $\pm0.1$ dex.  For further discussion of systematic uncertainties, see \cite{Almaini_etal:2017}.

For our morphological analyses, we use the PRIMER-UDS \emph{JWST}/NIRCam imaging.  Approximately $12$ per cent of the UDS DR11 galaxy catalogue from \cite{Wilkinson_etal:2021} lies within the PRIMER-UDS field and therefore has available \emph{JWST}/NIRCam coverage.  This yields a final sample of $9485$ star-forming, $610$ passive, and $228$ PSB galaxies.  We use this subset of PRIMER-UDS galaxies for all morphological analyses presented in this study.

\subsection{Stellar mass distributions}

\label{Stellar mass distributions}

\begin{figure}
\includegraphics[width=0.475\textwidth]{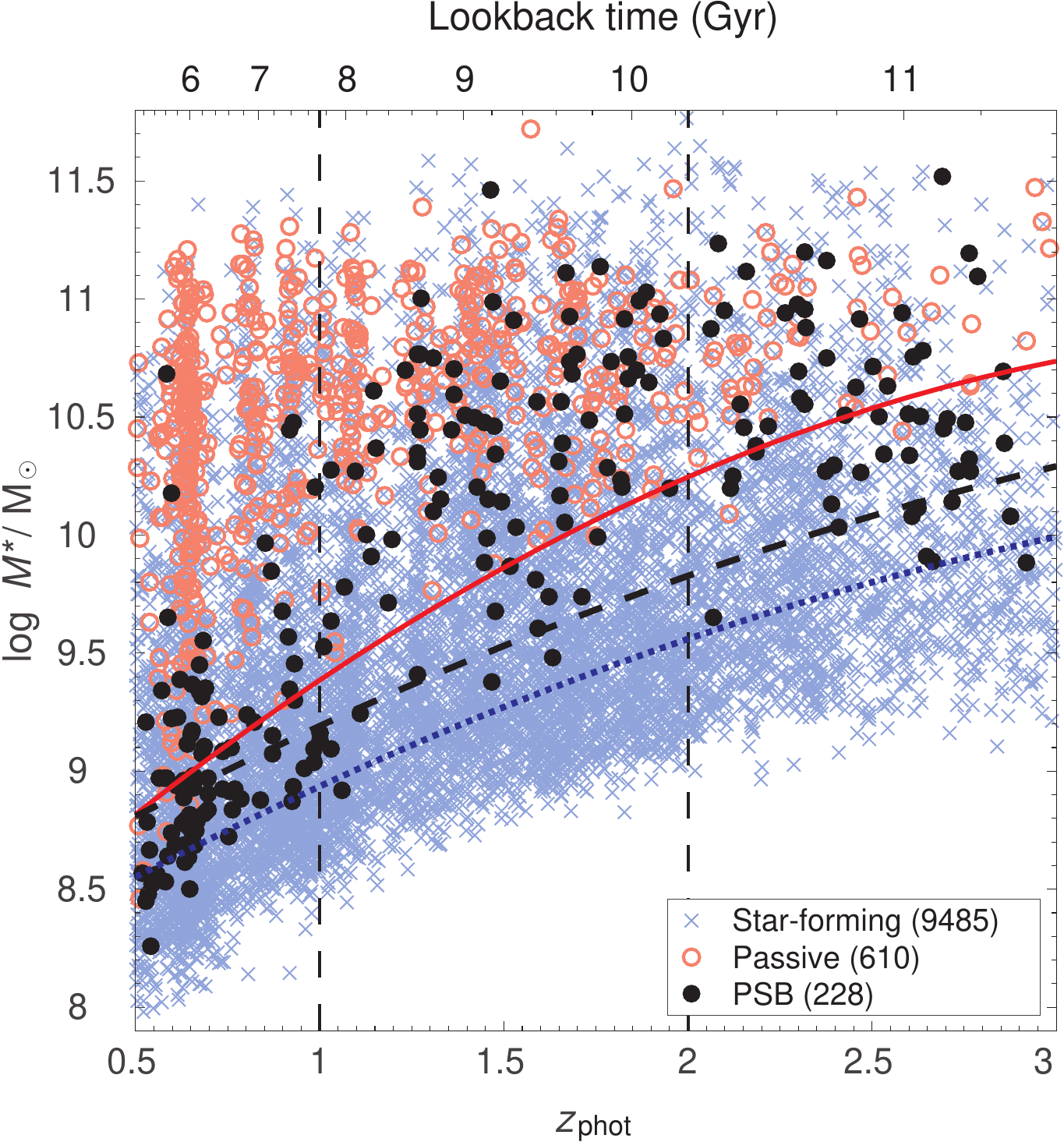}
\centering
\vspace{-0.0cm}
\caption{\label{mass-vs-z} The stellar mass $M_*$ distribution as a function of photometric redshift for star-forming (blue crosses), passive (red circles) and PSB galaxies (black points), within the PRIMER-UDS field.  Respective sample sizes are shown in the legend.  PSB galaxies exhibit significant evolution in their stellar mass distribution over $0.5 < z < 3$ (see e.g. \citealt{Wild_etal:2016, Taylor_etal:2023}).  In this paper, we take this evolution into account by considering the structural properties of these galaxies in three different epochs: low-$z$ ($0.5 < z < 1$), intermediate-$z$ ($1 < z < 2$) and high-$z$ ($2 < z < 3$).  These are separated by black dashed lines.  The $90$ per cent mass completeness curves for each population, star-forming (blue dotted line), passive (red line) and PSB (black dashed line), are also shown for reference.}
\end{figure}

In Fig.~\ref{mass-vs-z}, we present the stellar mass $M_*$ distribution as a function of redshift for star-forming, passive, and PSB galaxies within the PRIMER-UDS field.  Similarly to previous studies that use SC populations within the UDS \citep[e.g.][]{Wild_etal:2016, Maltby_etal:2018, Wilkinson_etal:2021}, we observe a strong evolution in the $M_*$ distribution of PSBs with redshift, in this case over $0.5 < z < 3$.  At $z < 1$, PSBs are predominantly low-mass ($M_* < 10^{\,9.5}\,\mathrm{M_{\odot}}$), whereas at $z > 1$, they are of higher mass ($M_* > 10^{\,9.5}\,\mathrm{M_\odot}$).  This marked trend suggests that high-mass PSBs at $z > 1$ represent a distinct population, with potentially different evolutionary histories and quenching mechanisms from their lower-redshift, low-mass counterparts.  We note that the absence of low-mass PSBs at high redshift ($z > 1$) may be intrinsic or could result from mass-completeness limitations.  To take this evolution into account, in this work we analyze the structural properties of PSBs within three redshift intervals: $0.5 < z < 1$, $1 < z < 2$, and $2 < z < 3$.

In this work, we use galaxy populations derived from the SC analysis of \cite{Wilkinson_etal:2021}, which was performed on UDS galaxies with $K_{\rm AB} < 24.5$ (see Section~\ref{Sample selection}).  Using this $K$-band limit, the equivalent mass completeness limits for each population were determined as a function of redshift using the method of \cite{Pozzetti_etal:2010}, applied to the full UDS DR11 sample (see Fig.~\ref{mass-vs-z}).  For each epoch and population, the $90$ per cent mass completeness limit at the upper end of the redshift bin is listed in Table~\ref{mass-comp-limits}.  However, when comparing the average global properties of the populations, we adopt the mass completeness limits of the PSB population -- the population of primary interest -- for all galaxy samples.  We note though, that adopting the stricter limits of the passive population has no significant effect on our results or conclusions.  The sizes of the resultant mass-complete samples are given in Table~\ref{galaxy-samples}.

\begin{table}
\centering
\begin{minipage}{82mm}
\centering
\caption{\label{mass-comp-limits} The 90 per cent mass completeness limits as determined for each galaxy population via the method of \citet{Pozzetti_etal:2010}. Limits are in units of ${\log}\,M_*/\rm{M}_\odot$ and determined at the upper limit of the respective redshift interval.  In this work, we adopt the mass completeness limits of the PSB
population for our galaxy samples.}
\begin{tabular}{lccccc}
\hline
{Sample}        &{Low-$z$}         &{}  &{Intermediate-$z$} &{} &{High-$z$}         \\
{}              &{$0.5 < z < 1$}   &{}  &{$1 < z < 2$}      &{}	&{$2 < z < 3$}      \\
\hline
Star-forming    &{$8.94$}          &{}  &{$9.56$}           &{}	&{$9.99$}           \\[0ex]
Passive         &{$9.39$}          &{}  &{$10.25$}          &{}	&{$10.74$}          \\[0ex]
PSB             &{$9.19$}          &{}  &{$9.83$}           &{}	&{$10.29$}          \\
\hline
\end{tabular}
\end{minipage}
\end{table}

\begin{table}
\centering
\begin{minipage}{82mm}
\centering
\caption{\label{galaxy-samples} The size of the mass-limited galaxy samples used throughout this work.  In this work, we adopt the 90 per cent mass completeness limits of the PSB population for our galaxy samples (see Table~\ref{mass-comp-limits}).}
\begin{tabular}{lccccc}
\hline
{Sample}        &{Low-$z$}         &{}  &{Intermediate-$z$} &{} &{High-$z$}         \\
{}              &{$0.5 < z < 1$}   &{}  &{$1 < z < 2$}		&{} &{$2 < z < 3$}      \\
\hline
Star-forming    &{$1079$}          &{}  &{$1304$}           &{} &{$306$}            \\[0ex]
Passive         &{$260$}           &{}  &{$224$}            &{} &{$47$}             \\[0ex]
PSB             &{$25$}            &{}  &{$60$}             &{} &{$34$}             \\
\hline
\end{tabular}
\end{minipage}
\end{table}

\section{Measuring structural parameters}

\label{Measuring structural parameters}

In this work, we use the eight-band \emph{JWST}/NIRCam imaging from PRIMER-UDS (see Section~\ref{PRIMER}) to determine the multiwavelength structure of our galaxy samples.  To measure galaxy structural parameters, we use the {\sc{galapagos}}{\scriptsize{-2}} data pipeline \citep{Haussler_etal:2022}, which automates source extraction and the fitting of S\'ersic models with {\sc{galfitm}} \citep{Haussler_etal:2013}.  While {{\sc{galapagos}}{\scriptsize{-2}}~/~{\sc{galfitm}} supports simultaneous multi-band S\'ersic fitting, leveraging all available multiwavelength data to produce a robust wavelength-dependent S\'ersic model, this means the structural parameters at each wavelength are not strictly independent.  However, because the aim of this study is to explore potential observational biases in earlier results based on single-band measurements \citep[e.g.][]{Almaini_etal:2017}, our analysis requires structural parameters that are independent at each wavelength.  To achieve this, we run {\sc{galapagos}}{\scriptsize{-2}} separately on each NIRCam waveband, yielding independently determined structural parameters across the rest-frame optical--near-infrared regime ($\lambda_{\rm obs} = 0.9$--$4.4\,\mu\mathrm{m}$).  A brief outline of the procedure is provided in the following sections.

\subsection{Source detection and sky subtraction}

\label{Source detection and sky subtraction}

The full PRIMER-UDS NIRCam mosaics measure $41\,800$ by $32\,000$ pixels, making full-frame processing with the {\sc{galapagos}}{\scriptsize{-2}} pipeline computationally prohibitive.  To overcome this, the mosaics were divided into $20$ sub-images arranged in a $4\times5$ grid, each approximately $8400\times8700$ pixels in size with a $200$-pixel overlap.  This overlap prevents sources near the edges from being split between sub-images and was applied uniformly to the science images as well as to the corresponding error and weight maps.  The {\sc{galapagos}}{\scriptsize{-2}} pipeline processes each sub-image individually, automatically removing any duplicate detections arising from overlapping regions during the construction of the final output catalogue.

For each sub-image, {\sc{galapagos}}{\scriptsize{-2}} performs a single {\sc{SExtractor}} \citep{Bertin&Arnouts:1996} run to identify sources and estimate their initial fitting parameters.  The {\sc{SExtractor}} parameters were tuned to provide an optimal extraction for the targets of interest, specifically high-redshift galaxies at $0.5 < z < 3$.   For each detected source, a postage stamp is extracted with dimensions equal to 2.5 times the {\sc SExtractor} Kron radius, which typically encloses more than $90$~per~cent of the source's total flux.  Corresponding postage stamps are also generated from the NIRCam error maps, which are used as noise maps during the {\sc galfitm} fitting process (see Section~\ref{structural parameters}).

Prior to S\'ersic modelling, {\sc{galapagos}}{\scriptsize{-2}} estimates the local sky background for each source.  Although the PRIMER–UDS NIRCam images are already sky-subtracted, some residual background can remain and affect the modelling of individual galaxies.  Therefore, an additional local background subtraction is applied on a galaxy--galaxy basis.   While {\sc{galfitm}} can set the sky as a free parameter in the fitting, the postage stamps used are often too small for a robust determination.  Consequently, {\sc{galapagos}}{\scriptsize{-2}} estimates the residual sky from a flux growth curve that uses the full science frame.  The resultant background is then incorporated as a fixed component in the subsequent {\sc{galfitm}} modelling (see Section~\ref{structural parameters}).  The sky estimation process relies on Kron ellipses from {\sc{SExtractor}} to define source-free regions.  However, very bright foreground stars can have Kron ellipses that cover large fractions of the image, significantly limiting the available area for sky estimation.  To mitigate this, such bright objects are masked during source detection, ensuring an accurate local sky background can be determined for each source.

\subsection{PSF determination}

\label{PSF determination}

Accurate determination of the point spread function (PSF) is essential for modelling galaxy light distributions, especially for distant galaxies where the PSF full-width half-maximum (FWHM) can approach the galaxy's angular size.  Even small mismatches in the PSF's extended structure can lead to significant biases in the derived structural parameters.  Although standard model and empirical PSFs are available for \emph{JWST}/NIRCam, the PSF is invariably affected by the image reduction process.  For this reason, it is best practice to construct high-quality empirical PSFs directly from the dataset used, in this case the PRIMER-UDS imaging, to ensure robust and reliable structural measurements.

To construct our \emph{JWST}/NIRCam PSFs, we use isolated stars previously identified in the UDS field \citep[e.g.][]{Lani_etal:2013, Almaini_etal:2017, Maltby_etal:2018} that lie within the PRIMER-UDS footprint.  For each filter, these stars are matched to sources extracted from the corresponding image using {\sc{SExtractor}}, yielding around $100$ stars per waveband.  Saturated stars, those with non-stellar profiles ($\texttt{class\_star} < 0.9$), and the faintest stars ($\texttt{magbest} > 22$) are excluded on a band-by-band basis.  Postage stamps of size $4\times4\,\mathrm{arcsec}^2$ are then created for each star and normalized to unit total flux.  Any nearby objects or artefacts within the postage stamps are carefully masked using dilated {\sc{SExtractor}} segmentation maps.  To ensure accurate alignment, the stars are resampled so their centroids coincide precisely with the centre of their postage stamp with sub-pixel accuracy.  Finally, the stamps are combined via a median stack to produce high signal-to-noise (S/N) empirical PSFs (see Fig.~\ref{psfs}).

\begin{figure*}
\includegraphics[width=0.8\textwidth]{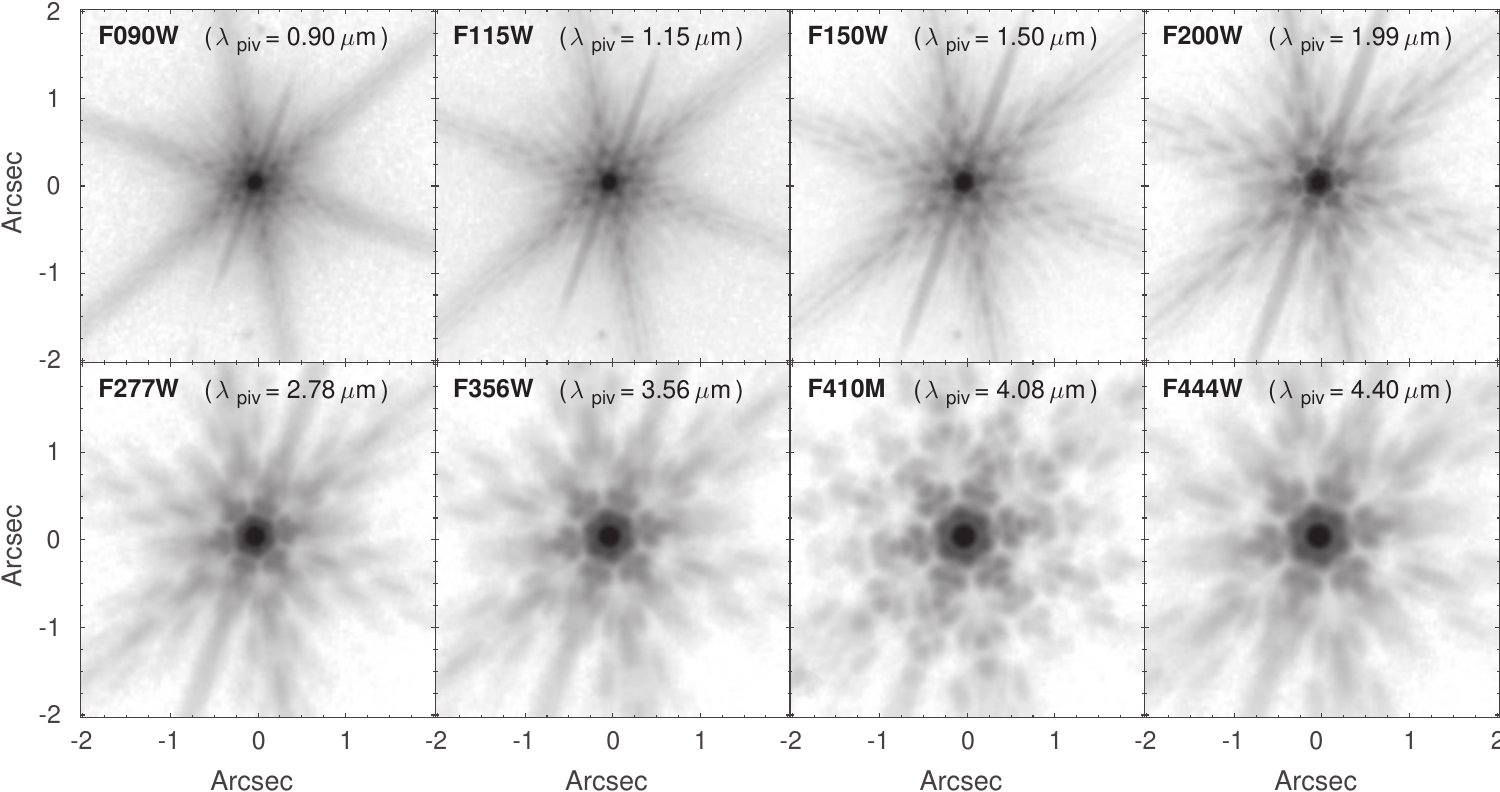}
\centering
\vspace{-0.0cm}
\caption{\label{psfs} Empirical \emph{JWST}/NIRCam PSFs for the PRIMER-UDS field.  PSFs for all eight PRIMER-UDS NIRCam filters are shown: $F090W$, $F115W$, $F150W$, $F200W$, $F277W$, $F356W$, $F410M$ and $F444W$.  Each PSF was constructed by stacking approximately $100$ isolated stars in the corresponding waveband.  The complex and extended structure of the NIRCam PSFs is clearly visible, including prominent diffraction features. A systematic increase in the FWHM with wavelength is observed, consistent with the diffraction-limited performance of \emph{JWST}.  The pivot wavelength ($\lambda_{\rm piv}$) for each filter is provided for reference.}
\end{figure*}

\begin{table}
\centering
\begin{minipage}{78mm}
\centering
\caption{\label{psf-fwhm} FWHM of the empirical \emph{JWST}/NIRCam PSFs determined from the PRIMER-UDS mosaics.  A systematic increase in the FWHM with wavelength is observed.  The pivot wavelength ($\lambda_{\rm piv}$) for each filter is provided for reference.}
\begin{tabular}{lcccccc}
\hline
\multicolumn{3}{c}{Short-Wavelength Channel}  &{}  &\multicolumn{3}{c}{Long-Wavelength Channel} \\[1ex]
Filter &{$\lambda_{\rm piv}$}  &FWHM          &{}  &Filter &{$\lambda_{\rm piv}$}  &FWHM        \\[0ex]
{}     &{($\mu\mathrm{m}$)}    &{}            &{}  &{}     &{($\mu\mathrm{m}$)}    &{}          \\[0ex]
\hline
$F090W$  &{$0.90$}  &{$0.068\arcsec$}         &{}  &$F277W$  &{$2.78$}  &{$0.126\arcsec$}       \\[0ex]
$F115W$  &{$1.15$}  &{$0.069\arcsec$}         &{}  &$F356W$  &{$3.56$}  &{$0.144\arcsec$}       \\[0ex]
$F150W$  &{$1.50$}  &{$0.074\arcsec$}         &{}  &$F140M$  &{$4.08$}  &{$0.157\arcsec$}       \\[0ex]
$F200W$  &{$1.99$}  &{$0.081\arcsec$}         &{}  &$F444W$  &{$4.40$}  &{$0.167\arcsec$}       \\
\hline
\end{tabular}
\end{minipage}
\end{table}

These NIRCam PSFs exhibit complex and extended structures that vary significantly between filters.  A systematic increase in the FWHM with observed wavelength is also evident, consistent with the diffraction-limited performance of \emph{JWST} (see Table~\ref{psf-fwhm}). 

In constructing these PSFs, particular attention was paid to the wings beyond the central FWHM, where even small inaccuracies can bias measurements of the structural parameters.  Our tests show that subtle mismatches in the extended wings can introduce systematic biases of $10$--$20$ per cent in the fitted parameters.  This highlights the importance of constructing high-fidelity empirical PSFs independently for each NIRCam image, on a case-by-case basis.


\begin{figure}
\includegraphics[width=0.475\textwidth]{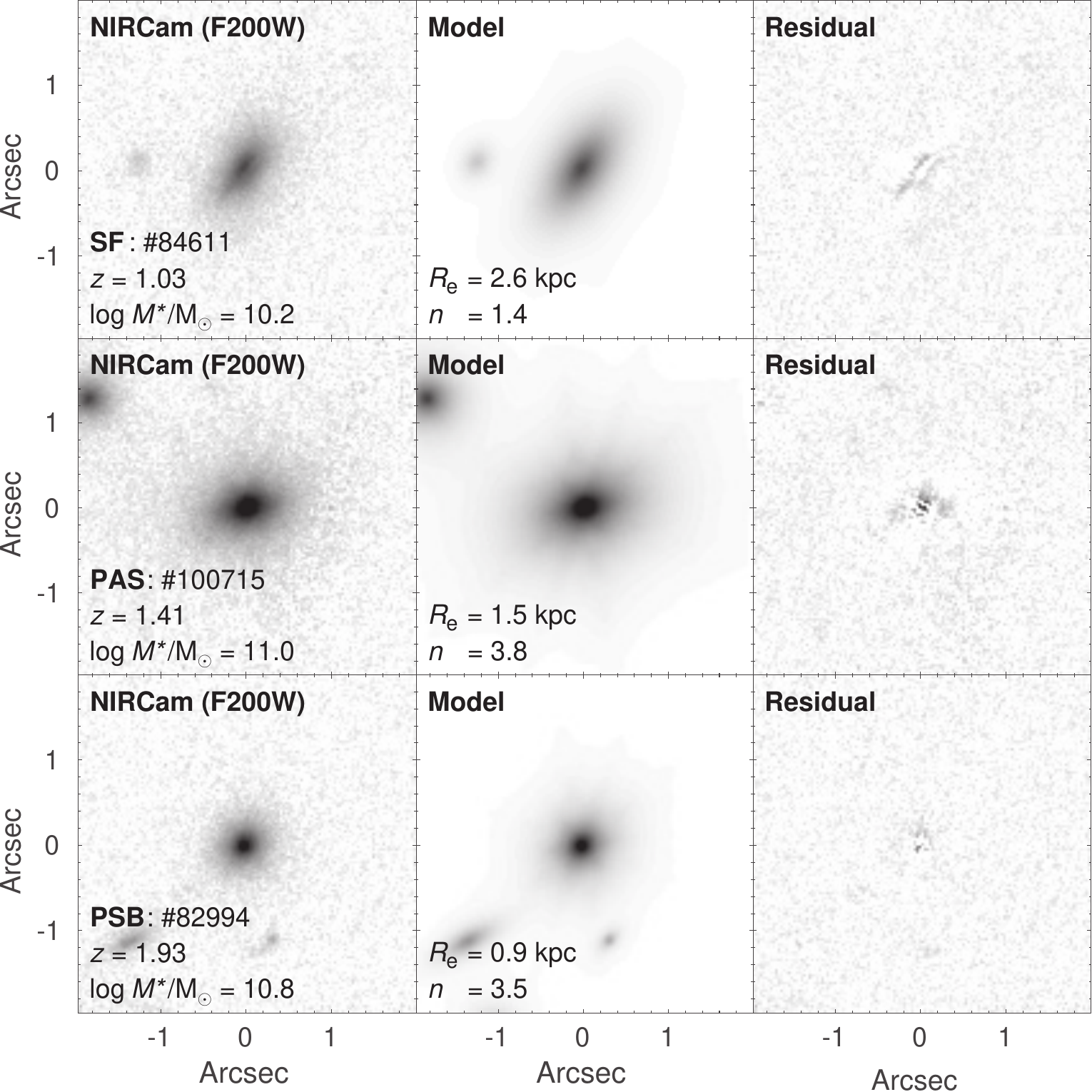}
\centering
\vspace{-0.0cm}
\caption{\label{example-fits} An example single S\'ersic fit for a typical $z > 1$ galaxy from each population: star-forming (top row), passive (middle row), and PSB (bottom row).  In each case, we display the $F200W$ science image (left-hand panels), the two-dimensional S\'ersic model from {\sc galfitm} (centre panels) and the residual image (right-hand panels).  In all cases, the galaxies are well-described by a single two-dimensional S\'ersic profile.}
\end{figure}

\begin{figure*}
\includegraphics[width=0.995\textwidth]{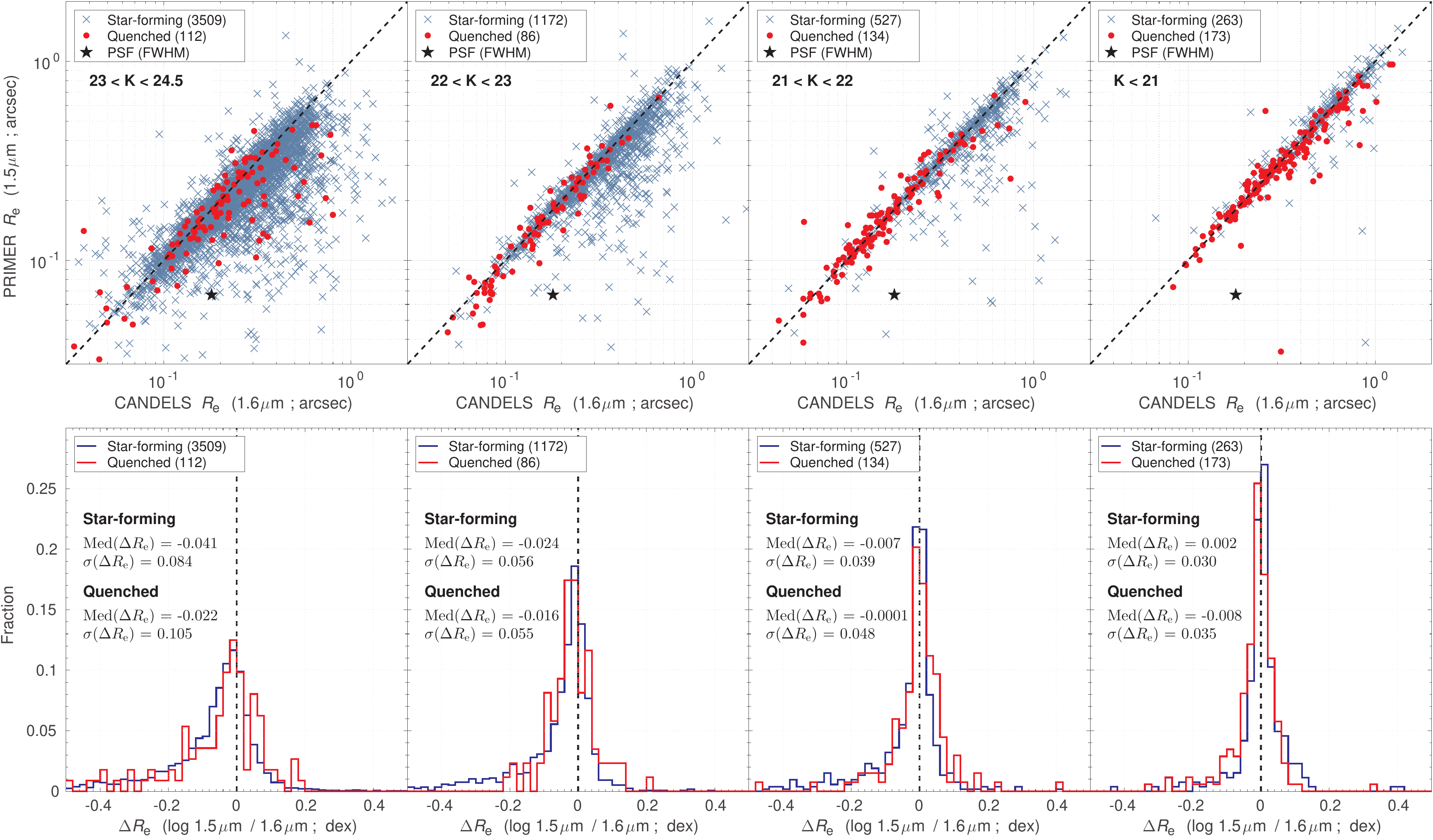}
\centering
\vspace{0.0cm}
\caption{\label{CANDELS-re-comp} A comparison of galaxy size ($R_{\rm e}$) measurements from PRIMER-UDS \emph{JWST}/NIRCam $F150W$ imaging ($1.5\rm\,\mu$m; this work) and CANDELS-UDS \emph{HST}/WFC3 $F160W$ imaging ($1.6\rm\,\mu$m; \citealt{vanderWel_etal:2012}).  Comparisons are shown for both star-forming and quenched galaxy populations, where \emph{quenched} refers to the combined passive and PSB populations.  \emph{Top row}: galaxy size ($R_{\rm e}$) comparisons across different $K$-band magnitude ranges.  The black dashed line indicates the one-to-one relation and the PSF FWHM is also shown for reference (black star).  \emph{Bottom row}: the difference ($\Delta R_{\rm e}$) between the PRIMER $1.5\rm\,\mu$m and CANDELS $1.6\,\mu$m size measurements across different $K$-band magnitude ranges, where $\Delta R_{\rm e} = {\rm log}\, [R_{\rm e}({\rm PRIMER}) / R_{\rm e}({\rm CANDELS})]$.  Respective sample sizes are shown in the legends.  In each case, the median and scatter in $\Delta R_{\rm e}$ -- Med($\Delta R_{\rm e}$) and $\sigma(\Delta R_{\rm e}$), respectively -- demonstrate there is very good agreement between the PRIMER and CANDELS sizes for both star-forming and quenched populations, across all magnitudes.}
\end{figure*}

\begin{figure*}
\includegraphics[width=0.995\textwidth]{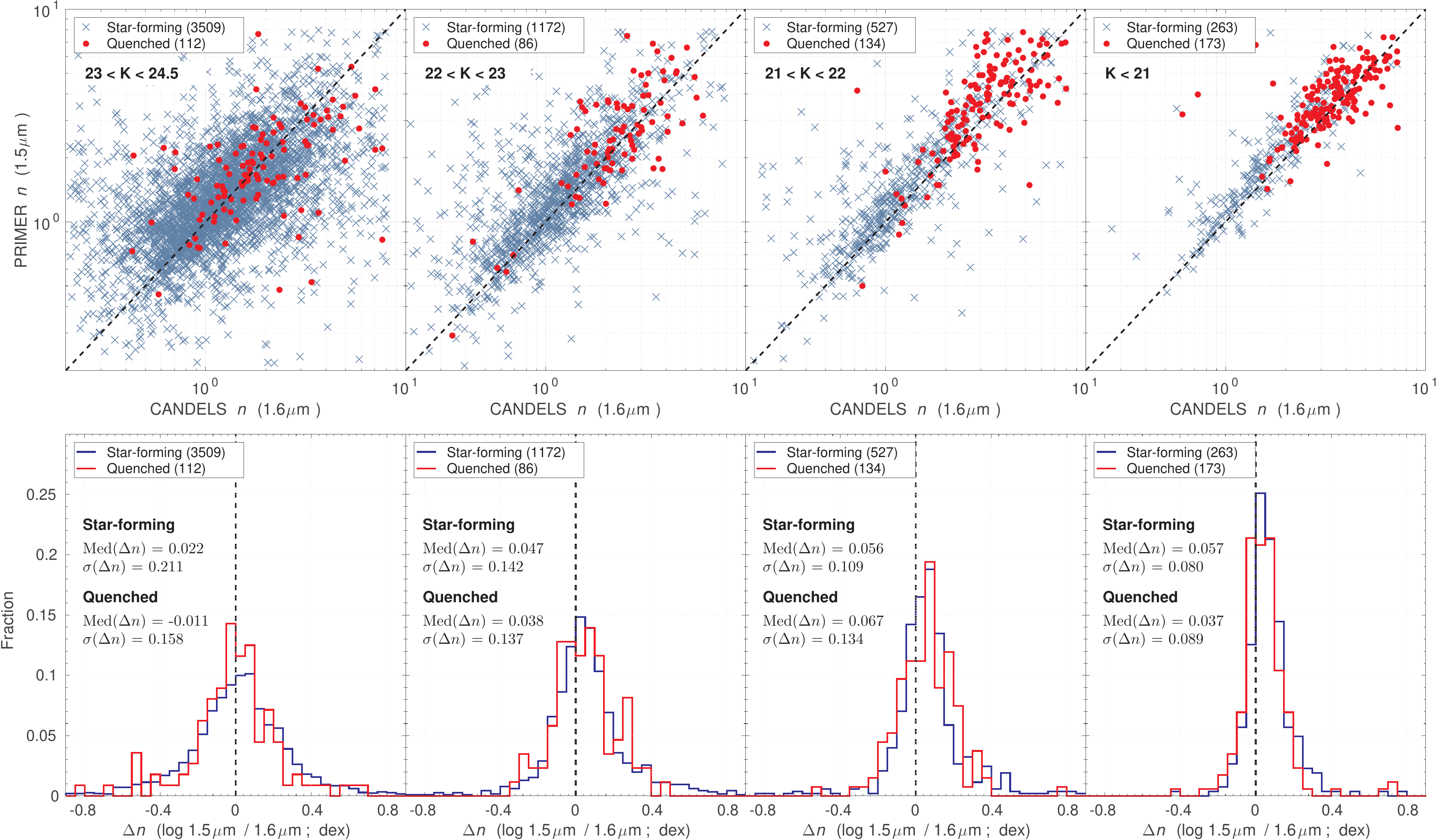}
\centering
\vspace{0.0cm}
\caption{\label{CANDELS-n-comp} A comparison of galaxy S\'ersic index ($n$) measurements from PRIMER-UDS \emph{JWST}/NIRCam $F150W$ imaging ($1.5\rm\,\mu$m; this work) and CANDELS-UDS \emph{HST}/WFC3 $F160W$ imaging ($1.6\rm\,\mu$m; \citealt{vanderWel_etal:2012}).  Comparisons are shown for both star-forming and quenched galaxy populations, where \emph{quenched} refers to the combined passive and PSB populations.  \emph{Top row}: galaxy S\'ersic index ($n$) comparisons across different $K$-band magnitude ranges.  The black dashed line indicates the one-to-one relation for reference.  \emph{Bottom row}: the difference ($\Delta n$) between the PRIMER $1.5\rm\,\mu$m and CANDELS $1.6\,\mu$m $n$~measurements across different $K$-band magnitude ranges, where $\Delta n = {\rm log}\, [n({\rm PRIMER}) / n({\rm CANDELS})]$.  Respective sample sizes are shown in the legends.  In each case, the median and scatter in $\Delta n$ -- Med($\Delta n$) and $\sigma(\Delta n)$, respectively -- demonstrate there is very good agreement between the PRIMER and CANDELS $n$ measurements for both star-forming and quenched populations, across all magnitudes.}
\end{figure*}

\subsection{Structural parameters}

\label{structural parameters}

To determine structural parameters, {\sc{galapagos}}{\scriptsize{-2}} employs {\sc{galfitm}} to fit S\'ersic models to the two-dimensional light distribution of each detected source.  In summary, PSF-convolved S\'ersic models are fit to the source light distribution, weighted by an appropriate error map, using a $\chi^2$ minimization \citep[see][]{ Peng_etal:2002, Haussler_etal:2013}.  The galaxies are modelled using a single-component S\'ersic profile of the form:
\begin{equation}
I(r) = I_{\rm e} \exp \left\{ -b_n \left[ \left( \frac{r}{R_{\rm e}} \right)^{1/n} - 1 \right] \right\},
\end{equation}
where $I_{\rm e}$ is the intensity at the effective radius $R_{\rm e}$, $n$ is the S\'ersic index, and $b_n$ is a normalization constant dependent on $n$.  In addition, the two-dimensional  S\'ersic model includes several other free parameters: the $(x, y)$ position of the centroid, the axis ratio $q$, and the position angle PA.  The base-level offset, i.e.\ the local sky background, is fixed at the value determined for the galaxy by the {\sc{galapagos}}{\scriptsize{-2}} sky measurement routine (see Section~\ref{Source detection and sky subtraction}).  Regarding the S\'ersic model, we note that many high-redshift galaxies can exhibit more complex morphologies \citep[see e.g.][]{Bruce_etal:2014a}.  However, single S\'ersic fits offer a straightforward parameterization that facilitates the comparison of galaxy structure across different populations.

For each sub-image, \sc{galapagos}}{\scriptsize{-2}} processes the detected sources in a hierarchical fashion, using {\sc{galfitm}} to fit galaxies from brightest to faintest.  Neighbouring sources are either masked, substracted, or fit simultaneously depending on their relative brightness and proximity.  For faint galaxies, if nearby brighter sources have already been fit, then their best-fit models are subtracted from the image prior to fitting the subject.  This approach helps reduce contamination from bright neighbours and ensures robust structural measurements for the faintest galaxies \citep[see][for further details]{Haussler_etal:2022}.

For each PRIMER-UDS NIRCam filter, {\sc{galapagos}}{\scriptsize{-2}} was used to generate an independent catalogue of structural parameters for the sources detected in that waveband.  To construct a coherent multiwavelength dataset, these individual catalogues were cross-matched using the $F200W$ detections as the reference.  This approach yields a consistent set of structural measurements across all eight NIRCam wavebands for sources identified in $F200W$.  The $F200W$ filter was selected as the reference because it is the NIRCam equivalent of the $K$-band, providing optimal correspondence with the $K$-band selected galaxy sample from the UDS used in this work (see Section~\ref{Sample selection}).

Fig.~\ref{example-fits} presents example fits in the $F200W$ waveband for representative $z > 1$ galaxies from each population: star-forming, passive and PSB.  The selected galaxies have effective radii ($R_{\rm e}$) and S\'ersic indices $n$ close to the median values of their respective populations.  For each galaxy, we display the science image, the corresponding {\sc{galfitm}} model, and the residual image (science minus model).  These examples demonstrate that, in general, galaxies from each population are well described by a single two-dimensional S\'ersic profile.

In the subsequent analysis, for each NIRCam filter, we exclude galaxies for which {\sc{galfitm}} either failed to fit or did not converge on a solution (e.g.\ due to the galaxy being too faint or a fitting constraint being reached).  Table~\ref{failed-fits} summarizes the fraction of galaxies rejected from our samples across the different NIRCam filters and galaxy populations.  The rejection rate typically remains below $10$ per cent, but is noticeably higher in the bluest NIRCam filter ($F090W$).  This trend reflects the fact that our galaxies are generally significantly fainter in this waveband compared to those at longer wavelengths.  Additionally, across all filters, the rejection rate is marginally higher for quenched populations (passive and PSB) than for star-forming galaxies.  This is likely a direct consequence of the typically more compact nature of quenched galaxies, which increases the incidence of unstable fits with {\sc{galfitm}}.

\begin{table}
\centering
\begin{minipage}{65mm}
\centering
\caption{\label{failed-fits} The fraction of galaxies rejected from our samples due to failed or constrained (i.e.\ non-convergent) {\sc{galfitm}} fits.  Values are for each population and NIRCam filter.}
\begin{tabular}{lccccc}
\hline
{Filter}     &{Star-forming} &{Passive}      &{PSB}          &{Total}        \\
\hline
$F$090$W$    &{$0.134$}      &{$0.156$}      &{$0.162$}      &{$0.136$} \\
$F$115$W$    &{$0.084$}      &{$0.118$}      &{$0.110$}      &{$0.086$} \\
$F$150$W$    &{$0.064$}      &{$0.115$}      &{$0.118$}      &{$0.068$} \\
$F$200$W$    &{$0.052$}      &{$0.095$}      &{$0.118$}      &{$0.056$} \\
$F$277$W$    &{$0.061$}      &{$0.087$}      &{$0.092$}      &{$0.063$} \\
$F$356$W$    &{$0.059$}      &{$0.087$}      &{$0.092$}      &{$0.062$} \\
$F$410$W$    &{$0.059$}      &{$0.090$}      &{$0.070$}      &{$0.061$} \\
$F$444$W$    &{$0.064$}      &{$0.085$}      &{$0.097$}      &{$0.066$} \\
\hline
\end{tabular}
\end{minipage}
\end{table}

\subsection{Comparison with previous works (CANDELS-UDS)}

\label{Comparison with previous works}

For the PRIMER-UDS field, galaxy structural parameters derived from complementary CANDELS \emph{HST} $J$- and $H$-band imaging ($F125W$ and $F160W$, respectively) are available from \cite{vanderWel_etal:2012}.  These measurements of size ($R_{\rm e}$) and S\'ersic index ($n$) cover approximately $65$ per cent of our galaxy sample in PRIMER-UDS.  Therefore, to assess the robustness of our \emph{JWST} NIRCam-derived structural parameters ($R_{\rm e}$ and $n$), we compare our $F150W$ ($1.5\,\mu\mathrm{m}$) measurements to those obtained from the CANDELS \emph{HST} $F160W$ imaging ($1.6\,\mu\mathrm{m}$).  While some differences are expected due to variations in imaging depth, PSF size, and fitting methodology, for a given galaxy population, the structural parameters should remain broadly consistent given the small difference in observed wavelength.

In Fig.~\ref{CANDELS-re-comp}, we present the comparison of galaxy size ($R_{\rm e}$) between PRIMER and CANDELS.  Comparisons are shown for star-forming and quenched galaxy populations across different $K$-band magnitude ranges, where \emph{quenched} refers to the combined passive and PSB populations.  Across all magnitude bins, we find good agreement between the PRIMER and CANDELS size measurements, with a tight correlation that becomes progressively stronger for brighter galaxies and for quenched systems.  We quantify the difference between the datasets as $\Delta R_{\rm e} = \log\left[R_{\rm e}({\rm PRIMER}) / R_{\rm e}({\rm CANDELS})\right]$, and the corresponding histograms further illustrate this consistency.  The median and scatter in $\Delta R_{\rm e}$ are small for both star-forming and quenched galaxies, indicating overall consistency between the datasets.  As expected, both the offset and scatter increase modestly toward fainter magnitudes, likely due to decreasing S/N and greater uncertainty in the structural fits.  Nonetheless, the agreement remains robust, with a median offset of less than $10$ per cent even at the faintest magnitudes.  Furthermore, we find galaxy sizes remain consistent even below the PSF FWHM (black star) of the respective imaging, demonstrating both the robustness of the recovered structural parameters and the effectiveness of the PSF-convolved modeling techniques employed in both datasets.  At the faintest magnitudes, the slight tail toward negative $\Delta R_{\rm e}$ likely reflects cases where the greater depth and higher spatial resolution of the PRIMER {\em JWST} imaging enables more accurate recovery of compact light profiles compared to the CANDELS {\em HST} imaging.  This effect is more pronounced for the faintest, low surface brightness sources, resulting in smaller size measurements for a subset of the faintest galaxies.

Similarly, Fig.~\ref{CANDELS-n-comp} presents analogous comparisons of the S\'ersic index ($n$) measurements between PRIMER and CANDELS.  As with galaxy size, the PRIMER and CANDELS measurements are in close agreement, exhibiting a correlation that strengthens towards brighter magnitudes.  We quantify the difference between the datasets as $\Delta n = \log\left[n({\rm PRIMER}) / n({\rm CANDELS})\right]$, and the corresponding histograms clearly highlight this consistency.  Due to the nature of the S\'ersic parameter, the scatter is higher than that observed for galaxy size ($R_{\rm e}$).  Nonetheless, the results indicate that structural interpretations based on S\'ersic index (e.g.\ distinguishing between disc- and spheroid-dominated systems) remain robust between the two datasets, even down to faint magnitudes.

Taken together, these comparisons validate the reliability of the structural parameters derived from the PRIMER \emph{JWST}/NIRCam imaging.  The consistency observed across both star-forming and quenched populations, and over a wide range of magnitudes, indicates that our structural parameters are in strong agreement with those of previous determinations.

For completeness, we also compared our $F200W$ NIRCam effective radii with those derived from the ground-based UDS $K$-band imaging \citep{Almaini_etal:2017}.  Although the significantly broader PSF ($\textrm{FWHM}\sim0.7\arcsec$) and lower resolution of the ground-based data limit its utility as a stringent validation test, we nevertheless find strong agreement, including for galaxies with intrinsic sizes well below the FWHM of the UDS $K$-band PSF (see Appendix~\ref{Supplementary Figures and Tables}; Fig.~\ref{UDS-re-comp}).  This consistency provides an important validation of the quality of the UDS $K$-band structural catalogue, which spans a far larger field ($\sim0.8\rm\,deg^2$) than the PRIMER-UDS footprint.

\begin{figure*}
\includegraphics[width=0.90\textwidth]{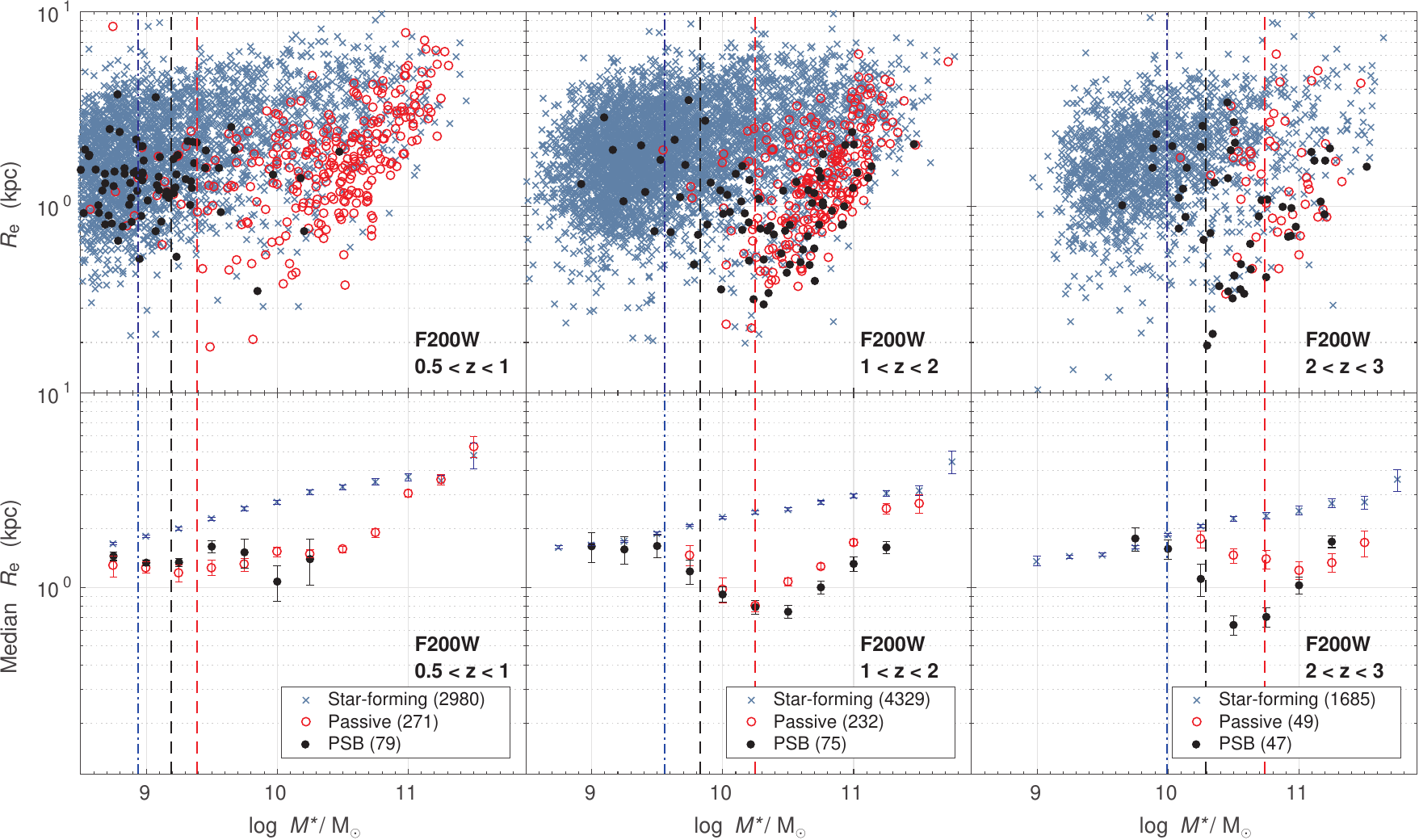}
\centering
\vspace{0.1cm}
\caption{\label{mass-size-f200w} The evolution of the stellar-mass--size relation for different populations using sizes ($R_{\rm e}$) determined at $2\rm\,\mu$m ($F200W$).  \emph{Top~row}: stellar mass vs.\ $R_{\rm e}$ for individual galaxies across three redshift intervals: $0.5 < z < 1$ (left-hand panel), $1 < z < 2$ (centre panel) and $2 < z < 3$ (right-hand panel).  \emph{Bottom row}: the corresponding median $R_{\rm e}$ as a function of stellar mass for each population and redshift interval.  Median $R_{\rm e}$ values are computed in stellar mass bins of $0.25$~dex, and only shown for bins containing more than two galaxies.  Respective sample sizes are shown in the legend.  The vertical lines represent the 90 per cent mass completeness limits for each population, determined at the upper limit of the respective redshift interval (see Table~\ref{mass-comp-limits}): star-forming (blue dot-dashed), passive (red dashed) and PSB (black dashed).  Errors in the median $R_{\rm e}$ ($1\sigma$) represent the standard error on the median value.  For both star-forming and passive galaxy populations, the expected size evolution is clearly observed: at fixed stellar mass galaxy size increases with decreasing redshift.}
\end{figure*}

\section{The structure of post-starburst galaxies}

\label{The structure of post-starburst galaxies}

In this section, we compare the structural properties of various galaxy populations (star-forming, passive, and PSB) across the redshift range $0.5 < z < 3$.  We investigate how these structural characteristics evolve with cosmic time and how they vary with wavelength across the rest-frame optical--near-infrared regime.  Specifically, we examine the stellar distributions of these galaxies by comparing their physical sizes ($R_{\rm e}$) in Section~\ref{The sizes of post-starburst galaxies} and their S\'ersic indices ($n$) in Section~\ref{The Sersic indices of post-starburst galaxies}.  Finally, in Section~\ref{Multiwavelength structure}, we assess how the average values of both $R_{\rm e}$ and $n$ vary with wavelength, providing a multiwavelength perspective on the structural properties of these galaxies.

\subsection{The sizes of post-starburst galaxies}

\label{The sizes of post-starburst galaxies}

In Fig.~\ref{mass-size-f200w}, we compare the stellar-mass--size relations for star-forming, passive, and PSB galaxies across three redshift intervals: $0.5 < z < 1$, $1 < z < 2$, and $2 < z < 3$.  These relations use galaxy sizes ($R_{\rm e}$) measured at $2\,\mu\mathrm{m}$ ($F$200$W$), showing both individual galaxies and median sizes calculated in stellar mass bins of $0.25$ dex.  For each population and redshift interval, we also indicate the $90$ per cent mass completeness limits, determined at the upper bound of the respective redshift interval (see Table~\ref{mass-comp-limits}).  Uncertainties on the median $R_{\rm e}$ ($1\sigma$) correspond to the standard error on the median value.

Across the redshift range $0.5 < z < 3$, the mass--size relations of star-forming and passive galaxies exhibit the expected, well-established trends consistent with previous studies \citep[e.g.][]{Shen_etal:2003, Daddi_etal:2005, Trujillo_etal:2006, Trujillo_etal:2007, vanDokkum_etal:2008, vanderWel_etal:2014}.  At all epochs, passive galaxies are significantly more compact than star-forming galaxies of the same stellar mass.  Furthermore, passive galaxies display a steeper mass-size relation above the characteristic pivot mass ($M_* > 10^{10.5}\,\mathrm{M_\odot}$), which roughly corresponds to the mass scale separating disc- and spheroid-dominated systems.  Both populations also show the expected size evolution, with galaxy sizes increasing at fixed stellar mass toward lower redshifts.  Notably, passive galaxies above the pivot mass increase in size at approximately twice the rate of star-forming galaxies, consistent with expectations \citep[see e.g.][]{Trujillo_etal:2007, vanderWel_etal:2014}.

For PSB galaxies, the mass--size relations across the redshift range $0.5 < z < 2$ also exhibit the well-established trends reported in previous studies \citep[e.g.][]{Whitaker_etal:2012a, Almaini_etal:2017, Maltby_etal:2018, Zhang_etal:2024}.  In the redshift interval $0.5 < z < 1$, PSBs are typically low-mass ($M_* < 10^{9.5}\,\mathrm{M_{\odot}}$) and, at fixed mass, are significantly more compact than star-forming galaxies, yet similar in size to low-mass passive galaxies.  In contrast, at higher redshift ($1 < z < 2$), PSBs are generally more massive ($M_* > 10^{9.5}\,\mathrm{M_{\odot}}$).  Above the pivot mass ($M_* > 10^{10.5}\,\mathrm{M_\odot}$), they are extremely compact, representing the most compact population at this epoch.  On average, these PSBs are $\sim25$ per cent smaller than passive galaxies of similar mass.  These results are consistent with the findings of \cite{Almaini_etal:2017} and \cite{Maltby_etal:2018}, who analyzed PSB structure in the UDS field using structural parameters derived from the ground-based UDS $K$-band imaging and CANDELS \emph{HST} $J$- and $H$-band data, respectively.

Interestingly, for PSBs at $1 < z < 2$, we also observe a distinct upturn in the mass--size relation toward lower masses, with PSBs below the pivot mass ($M_* < 10^{10.5}\,\mathrm{M_\odot}$) exhibiting a noticeable increase in size compared to their higher-mass counterparts.  This suggests that low-mass PSBs at these redshifts are less compact than expected based on the extrapolation of the high-mass trend, and points to the existence of a distinct PSB population at low masses, analogous to that observed at lower redshift.  This may indicate that the quenching pathway responsible for the formation of low-mass PSBs at $z < 1$, thought to differ from that of the higher-mass PSBs at $z > 1$ \citep[see e.g.][]{Maltby_etal:2018}, is already in place at these higher redshifts.  While this upturn occurs close to the 90 per cent mass completeness limit, and may therefore be affected by selection biases and/or incompleteness, it potentially reveals a previously unrecognized population of more extended, low-mass PSBs at these redshifts.  This trend mirrors the flattening of the size–mass relation reported for low-mass quiescent galaxies at $z > 1$ in recent \emph{JWST} PRIMER studies \citep[e.g.][]{Cutler_etal:2024, Hamadouche_etal:2025}, which is also consistent with a low-mass quenching pathway operating at these epochs.  A more comprehensive analysis of this trend, and its implications for PSB structural evolution, will be presented in future work using deeper datasets (de Lisle et al., in preparation). 

For PSBs at $2 < z <3$, we find similar behaviour to that observed at $1 < z <2$.  PSBs above the pivot mass ($M_* > 10^{10.5}\,\mathrm{M_\odot}$) are extremely compact, and on average more compact than the passive population.  However, at the highest masses ($M_* > 10^{11}\,\mathrm{M_\odot}$), PSBs and passive galaxies exhibit comparable sizes.  We note, though, that at these redshifts the distinction between passive and PSB galaxies becomes less well defined, as the young age of the Universe ($\sim2\,\rm{Gyr}$ at $z\sim2$) implies that essentially all passive systems are likely to have been quenched in the recent past and on short timescales.

In addition to the $2\,\mu\mathrm{m}$ ($F$200$W$) measurements, we also examine the stellar-mass--size relation using sizes ($R_{\rm e}$) determined from seven other PRIMER-UDS \emph{JWST}/NIRCam wavebands ($F090W$, $F115W$, $F150W$, $F277W$, $F356W$, $F410M$, and $F444W$), spanning the near-infrared regime from $0.9$--$4.4\,\mu\mathrm{m}$.  An inspection of the mass--size relations for each galaxy population across these wavelengths reveals broadly consistent results with those discussed above, confirming the robustness of the structural trends identified at $2\,\mu\mathrm{m}$.  While we do not present all the relations here, we include the mass--size relations using sizes determined from the extremes of wavelength coverage for comparison (see Appendix~\ref{Supplementary Figures and Tables}).  These correspond to the $0.9\,\mu\mathrm{m}$ ($F090W$) and $4.4\,\mu\mathrm{m}$ ($F444W$) wavebands and are presented in Figs.~\ref{mass-size-f090w} and \ref{mass-size-f444w}, respectively.

We note that the use of a single observed-frame waveband (e.g.~$F$200$W$; $2\rm\,\mu{m}$) for our galaxy sizes results in a range of rest-frame wavelengths being probed across our redshift range.  Given that galaxy structural parameters can exhibit wavelength dependence, potentially reflecting different stellar populations being traced, this could affect comparisons between redshift bins.  We address this by constructing mass–size relations using the PRIMER waveband closest to a rest-frame wavelength of $0.8\rm\,\mu{m}$, and find no significant differences in the resulting trends (see Appendix~\ref{Supplementary Figures and Tables}; Fig.~\ref{mass-size-restframe}).  Consistent results are also obtained using rest-frame wavelengths of $1.2\rm\,\mu{m}$ and $1.6\rm\,\mu{m}$.  This demonstrates that our results are robust to wavelength-dependent effects and confirms that the use of $F$200$W$ as our fiducial band does not bias our conclusions.

\subsection{The S\'ersic indices of post-starburst galaxies}

\label{The Sersic indices of post-starburst galaxies}

\begin{figure}
\includegraphics[width=0.475\textwidth]{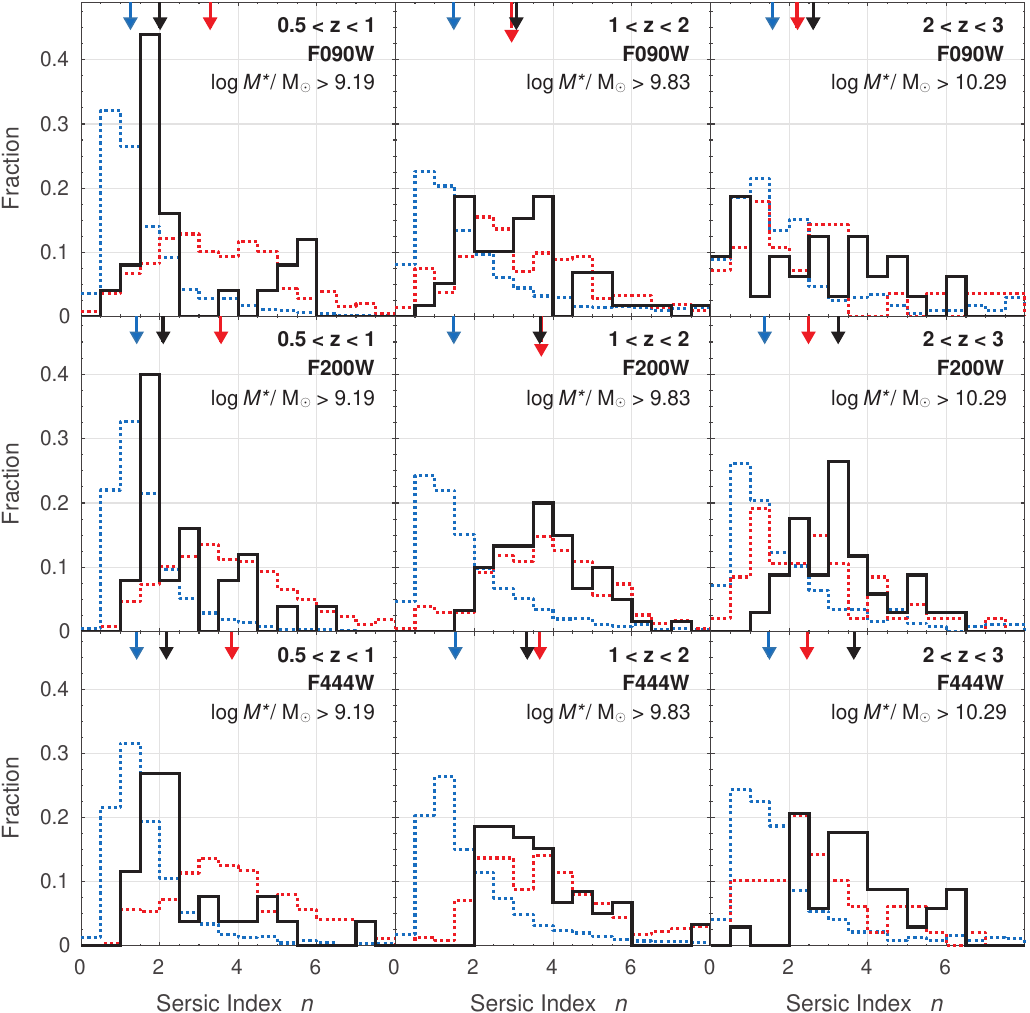}
\centering
\caption{\label{sersic-index} The S\'ersic index ($n$) distributions of star-forming (blue dotted), passive (red dotted) and PSB (black solid) galaxies across the three redshift intervals studied: $0.5 < z < 1$ (left-hand panels), $1 < z < 2$ (centre panels) and $2 < z < 3$ (right-hand panels).  Distributions are shown for $n$ values determined at $0.9\rm\,\mu$m ($F090W$; top row), $2\rm\,\mu$m ($F200W$; middle row) and $4.4\rm\,\mu$m ($F444W$; bottom row).  In each case, we apply the $90$ per cent mass-completeness limits of the PSB population, determined at the upper limit of the respective redshift interval (see Table~\ref{mass-comp-limits}).  Median values for each population are indicated by downward arrows matching their respective colours.}
\end{figure}

In Fig.~\ref{sersic-index} (middle row), we compare the $2\,\mu\mathrm{m}$ ($F$200$W$) S\'ersic index ($n$) distributions of star-forming, passive, and PSB galaxies in three redshift intervals: $0.5 < z < 1$, $1 < z < 2$, and $2 < z < 3$.  In each case, the distributions include only galaxies above the mass-completeness limit of the PSB population (see Table~\ref{mass-comp-limits}).  For star-forming galaxies, at all epochs, the S\'ersic index distributions peak at $n\sim1$.  This is indicative of a disc-dominated (generally rotationally supported) stellar distribution being typical for these galaxies, as expected.  For passive galaxies at $z < 2$, we find the S\'ersic index distributions peak at significantly higher values ($n\sim3$).  This is characteristic of a more spheroid-dominated stellar distribution, again as expected.  However, at $z > 2$, we find that massive passive galaxies exhibit lower S\'ersic indices compared to their lower redshift counterparts, with distributions more closely resembling those of the star-forming population (although still significantly different based on K-S tests; $p < 10^{-4}$).  This observation is consistent with other recent studies using structural parameters from \emph{JWST} \citep[e.g.][]{Ormerod_etal:2024} and suggests that a different quenching route is responsible for these $z > 2$ passive galaxies compared to their lower redshift counterparts.

For PSB galaxies, we observe a significant evolution in the S\'ersic index distribution with redshift.  At $z < 1$, the distribution differs significantly from both star-forming and passive populations (based on K-S tests; $p < 10^{-4}$), but more closely resembles that of star-forming galaxies, peaking at $n\sim2$.  This suggests that these recently quenched systems typically retain disc-dominated stellar structures that have somehow survived the quenching event.  In contrast, at $z > 1$, PSBs exhibit significantly higher S\'ersic indices than their star-forming progenitors, with distributions peaking at $n\sim3.5$.  Within the interval $1 < z < 2$, the PSB distribution is statistically indistinguishable from that of the massive passive population.  However, at $z > 2$, it differs significantly due to the passive population typically exhibiting lower S\'ersic indices (see above).  These results imply that, at $z > 1$, massive PSBs typically have spheroid-dominated stellar distributions, suggesting that their quenching involved a highly disruptive, dissipative event (e.g.\ major merger) that significantly altered their stellar distributions from their disc-dominated, star-forming progenitors.  Across the redshift range $0.5 < z < 3$, the evolution in the S\'ersic index distribution is likely linked to changes in the stellar mass distribution of PSBs (see Fig.~\ref{mass-vs-z}), and supports the interpretation that PSBs at different epochs (and/or mass regimes) are typically formed via different quenching mechanisms.  These findings are entirely consistent with previous results in the UDS field which use ground- and \emph{HST}-based structural parameters \citep{Almaini_etal:2017, Maltby_etal:2018} and extend them to $z > 2$ for the first time.

\begin{figure*}
\includegraphics[width=0.85\textwidth]{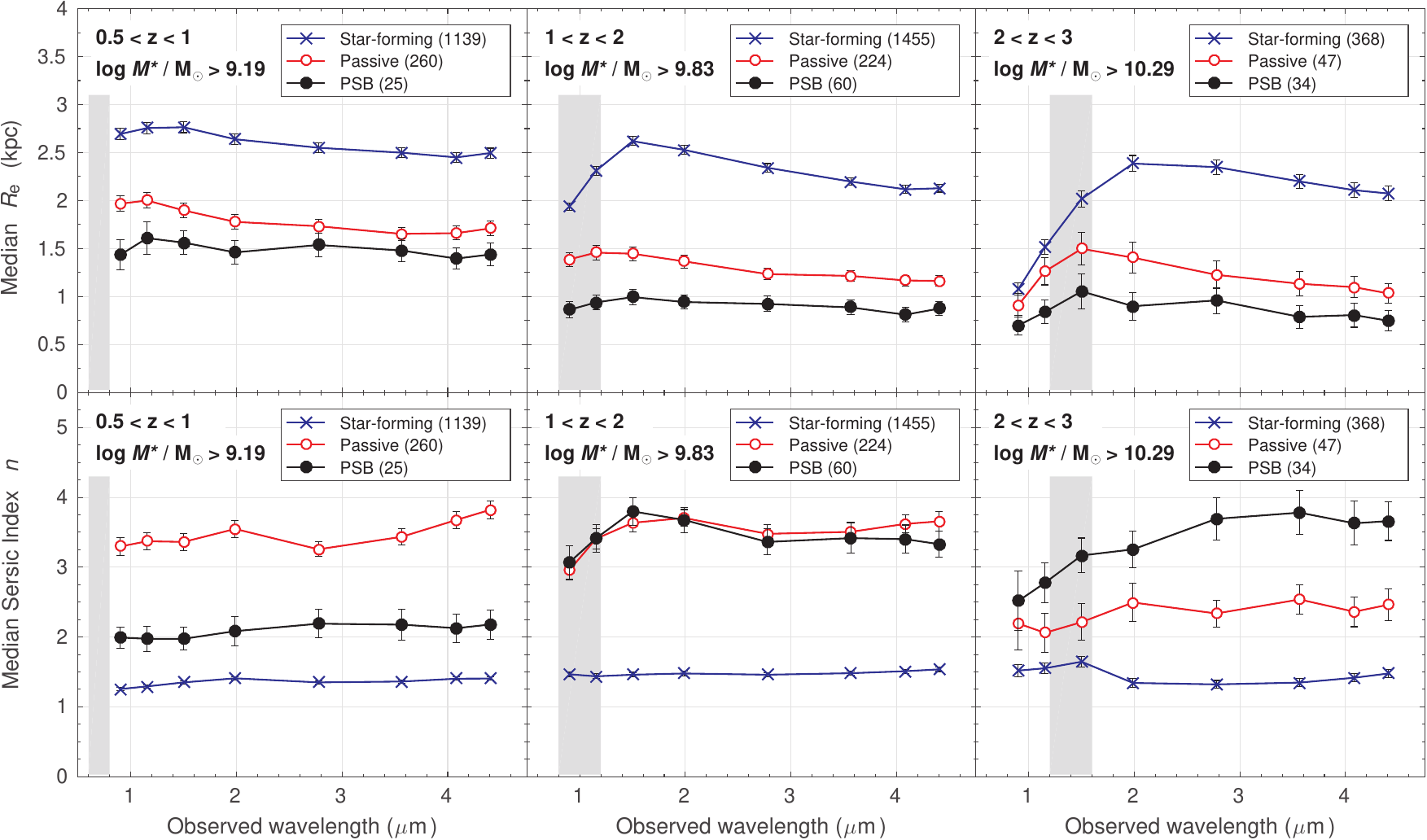}
\centering
\caption{\label{struct-vs-wave} The typical structure of different galaxy populations as a function of wavelength across the rest-frame optical--near-infrared regime ($\lambda_{\rm obs}\ 0.9$--$4.4\,\mu\rm{m}$).  \emph{Top row}: the median size ($R_{\rm e}$) as a function of wavelength for star-forming (blue crosses), passive (red circles) and PSB galaxies (black points) across three redshift intervals: $0.5 < z < 1$ (left-hand panels), $1 < z < 2$ (centre panels) and $2 < z < 3$ (right-hand panels).  \emph{Bottom row}: the median S\'ersic index ($n$) as a function of wavelength for the same galaxy populations.  Respective sample sizes are shown in the legend.  In each case, we apply the 90 per cent mass completeness limits of the PSB population (see Table~\ref{mass-comp-limits}).  Errors are the uncertainty ($1\sigma$) in the median as determined from $100$ bootstrapped simulations.  To indicate the rest-frame wavelengths being probed, shaded regions show the observed-frame range of the $4000\,$\AA\ break in each redshift bin.  Notably, at all epochs, PSBs exhibit remarkably consistent structural properties with wavelength.}
\end{figure*}

In addition to the $2\,\mu\mathrm{m}$ ($F$200$W$) measurements, we also examine the S\'ersic indices derived from seven other PRIMER-UDS \emph{JWST}/NIRCam wavebands ($F090W$, $F115W$, $F150W$, $F277W$, $F356W$, $F410M$, and $F444W$), spanning the near-infrared regime from $0.9$--$4.4\,\mu\mathrm{m}$.  An inspection of the S\'ersic index distributions for each galaxy population across these wavelengths reveals broadly consistent results with those discussed above, confirming the robustness of the structural trends identified at $2\,\mu\mathrm{m}$.  While we do not present all distributions here, we include the S\'ersic index distributions from the extremes of wavelength coverage for comparison (see Fig.~\ref{sersic-index}).  These correspond to the $0.9\,\mu\mathrm{m}$ ($F090W$) and $4.4\,\mu\mathrm{m}$ ($F444W$) bands.  We expand on this result and provide a discussion of its implications in Section~\ref{Multiwavelength structure}.

\subsection{Multiwavelength structure}

\label{Multiwavelength structure}

It is well established that the size and structure of galaxies can vary with wavelength.  This variation mainly arises from differences in the spatial distribution of the stellar populations probed, which may differ in age and chemical composition, as well as the effects of dust obscuration \citep[e.g.][]{Vulcani_etal:2014, Kennedy_etal:2015}.  The presence of age gradients and/or dust obscuration within galaxy populations can bias the determination of their structural parameters if measurements are based on data from a narrow wavelength range.  Such biases, if unaccounted for, could lead to incorrect interpretations of their evolutionary histories.  In this section, we test for such biases by using the structural parameters from all eight NIRCam filters to explore the multiwavelength structure of our galaxy populations.

In Fig.~\ref{struct-vs-wave}, we present the typical structural properties of our galaxy populations as a function of wavelength across the rest-frame optical--near-infrared regime ($\lambda_{\mathrm{obs}} = 0.9$--$4.4\,\mu\mathrm{m}$).  We compare the median size ($R_{\rm e}$) and S\'ersic index ($n$) as functions of observed wavelength for star-forming, passive, and PSB galaxies across the three epochs studied: $0.5 < z < 1$, $1 < z < 2$, and $2 < z < 3$.  The corresponding values are also summarized in Appendix~\ref{Supplementary Figures and Tables} (Table~\ref{struct-vs-wave-table}).  In each case, median values are based only on galaxies above the mass completeness limit of the PSB population (see Table~\ref{mass-comp-limits}).  However, we note that repeating the analysis using a range of higher stellar mass limits has no significant effect on our conclusions.  Uncertainties in the median values ($1\sigma$) are determined from the variance between medians derived from $100$ bootstrapped simulations.

At $0.5 < z < 1$, for each galaxy population, we find the structural parameters measured at $2\,\mu\mathrm{m}$ ($F200W$) remain broadly consistent across the wavelength range $0.9$--$4.4\,\mu\mathrm{m}$.  At all wavelengths, passive galaxies are significantly more compact than star-forming galaxies, and have high S\'ersic indices ($n\sim3.5$) characteristic of a spheroidal stellar distribution.  In contrast, star-forming galaxies consistently show low S\'ersic indices ($n\sim1.2$) across the entire wavelength range, indicative of a disc-dominated, generally rotationally supported structure.  For both passive and star-forming populations, although the structure is broadly consistent with wavelength, we do observe a slight increase in size at $\lambda_{\rm obs} < 2\,\mu\mathrm{m}$.  This trend may reflect the presence of younger stars in the outskirts of these galaxies, consistent with expectations from inside--out growth \citep[e.g.][]{Tacchella_etal:2016}, or the accretion of younger stars into an outer envelope through minor mergers \citep[e.g.][]{Naab_etal:2009}.

With respect to PSBs at $0.5 < z < 1$, there is remarkable consistency in their structural parameters across the wavelength range probed.  Regardless of wavelength, PSBs are compact ($R_{\rm e}\sim1.5\,\mathrm{kpc}$) and exhibit S\'ersic indices intermediate between the star-forming and passive populations ($n\sim2$), yet still characteristic of a disc-dominated system.  These PSBs are typically low-mass systems ($M_* < 10^{9.5}\,\mathrm{M_{\odot}}$; see Fig.~\ref{mass-vs-z}).  When limiting the star-forming and passive populations to the same mass regime, we find that PSBs at this epoch have structural properties similar to those of the low-mass passive population (i.e.\ passive discs), the population into which they will most likely evolve.  These findings are entirely consistent with those of \cite{Maltby_etal:2018} regarding PSB structure at $0.5 < z < 1$ and $\lambda_{\rm obs} < 1.6\,\mu\mathrm{m}$, while for the first time extending the analysis deep into the near-infrared, out to $\lambda_{\rm obs} < 4.4\,\mu\mathrm{m}$.  Our results demonstrate that the compact disc structure of PSBs persists well into the near-infrared, and is not biased by significant age-gradients and/or dust obscuration.

At $1 < z < 2$, for each galaxy population, the structural parameters measured at $2\,\mu\mathrm{m}$ ($F200W$) remain broadly constant across the wavelength range $0.9$--$4.4,\mu\mathrm{m}$.  The star-forming and passive populations exhibit trends similar to those observed at lower redshift, but with slightly smaller sizes, consistent with expected size evolution.  For star-forming galaxies, however, we observe a significant decrease in size at $\lambda_{\rm obs} < 1.5\,\mu\mathrm{m}$, corresponding to rest-frame wavelengths $\lambda_{\rm rest} \lessapprox 0.4\,\mu\mathrm{m}$ that trace younger stellar populations (i.e.\ O, B, A, F stars).  This decrease may result from centralized star formation and/or dust obscuration in the outer regions of these galaxies.

For PSBs at $1 < z < 2$, regardless of wavelength, the stellar distribution is extremely compact ($R_{\rm e}\sim1\,\rm kpc$) and characterized by a high S\'ersic index ($n\sim3.5$).  Their structures resemble those of the passive population but considerably more compact.  We find no evidence for significant colour gradients, implying a lack of notable age gradients and/or dust obscuration.  This suggests that previous findings based on structural parameters measured over a narrow wavelength range are robust \citep[e.g.][]{Almaini_etal:2017, Maltby_etal:2018}. A study by \citet[]{Suess_etal:2020} suggested that the more compact nature of PSBs at $z>1$ may be caused by colour gradients, since younger quiescent galaxies were found to show flatter colour gradients compared to older systems \citep[see also][]{Suess_etal:2022}. Our study does show some evidence that older quiescent galaxies are more compact at $4.4\,\mu\rm{m}$ compared to $\sim 1.5\,\mu\rm{m}$ (see Fig.~\ref{struct-vs-wave}), in  agreement with \citet[]{Suess_etal:2022}. However, the PSBs remain significantly more compact than older passive galaxies at all wavelengths. We conclude that the size differences are not strongly affected by colour gradients.

At higher redshifts ($2 < z < 3$), the structural parameters measured at $2\,\mu\mathrm{m}$ ($F200W$) generally provide a reliable representation of the galaxy structure across the wavelength range $0.9$--$4.4\,\mu\mathrm{m}$.  We observe a continuation of the trends seen at $1 < z < 2$, but with some notable differences.  For example, at all wavelengths passive galaxies are more compact than star-forming galaxies, but in contrast to lower redshifts, present lower S\'ersic indices more indicative of disc-dominated systems.  This indicates these galaxies must have experienced a different evolutionary history to similar mass passive galaxies at lower redshift, in order for their disc-dominated structure to persist post quenching.  For star-forming galaxies, we also observe the decrease in size towards bluer wavelengths, but this time at $\lambda_{\rm obs} < 2\,\mu\mathrm{m}$.  This shift reflects the increase in redshift and still corresponds to $\lambda_{\rm rest} \lessapprox 0.4\,\mu\mathrm{m}$.

\begin{figure*}
\includegraphics[width=0.85\textwidth]{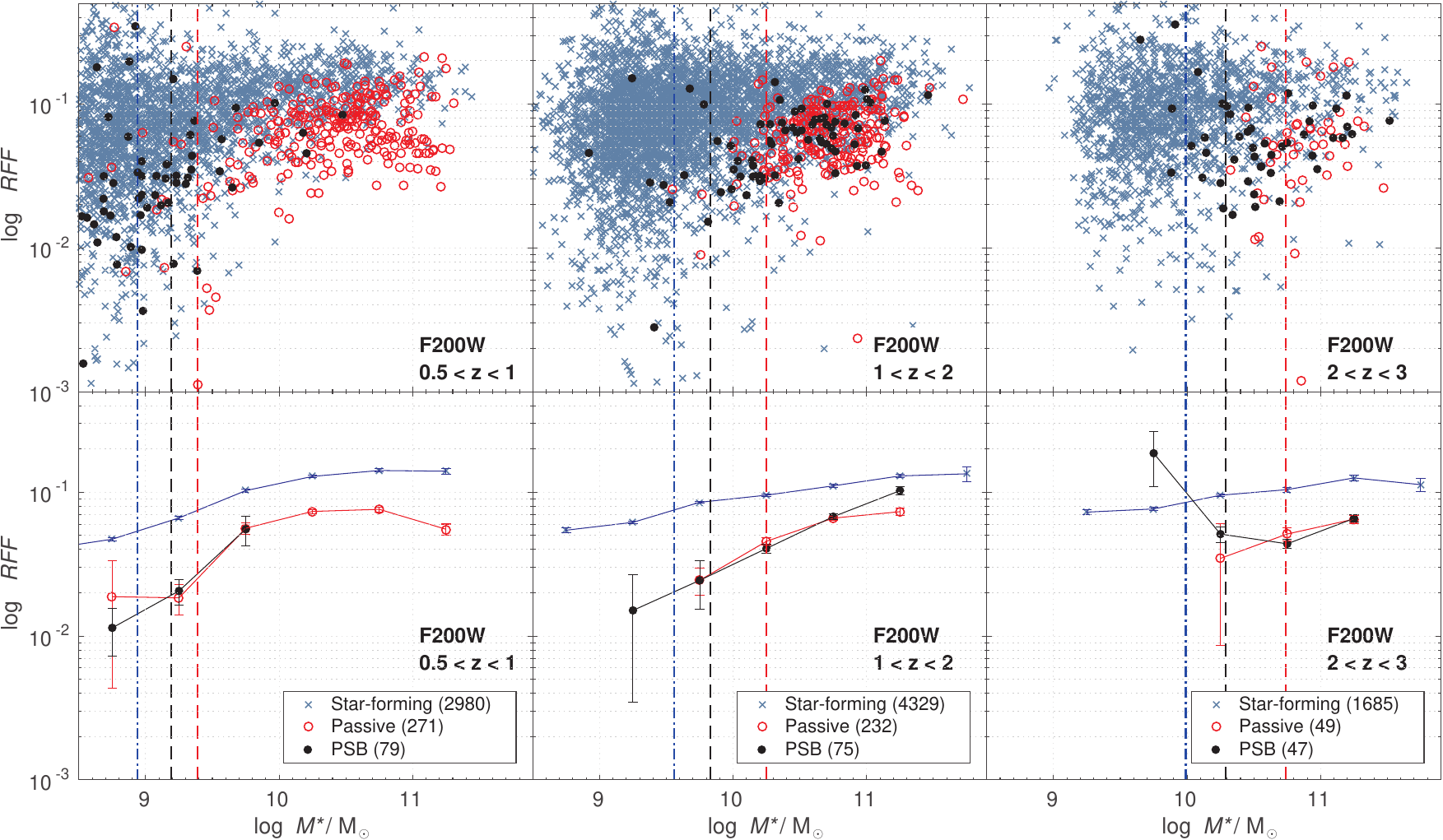}
\centering
\caption{\label{rff-vs-mass} The residual flux fraction ({\emph{RFF}}), determined at $2\rm\,\mu{m}$ ($F200W$), as a function of stellar mass for star-forming (blue crosses), passive (red circles) and PSB galaxies (black points).  \emph{Top row}: stellar mass vs.\ {\emph{RFF}} for individual galaxies across three redshift intervals: $0.5 < z < 1$ (left-hand panel), $1 < z < 2$ (centre panel) and $2 < z < 3$ (right-hand panel).  \emph{Bottom row}: the corresponding median {\emph{RFF}} as a function of stellar mass for each population and redshift interval.  Median {\emph{RFF}} values are computed in stellar mass bins of $0.5$~dex, and only shown for bins containing more than two galaxies.  Respective sample sizes are shown in the legend.  The vertical lines represent the 90 per cent mass completeness limits for each population, determined at the upper limit of the respective redshift interval: star-forming (blue dot-dashed), passive (red dashed) and PSB (black dashed).  Errors in the median {\emph{RFF}} ($1\sigma$) represent the standard error on the median value.}
\end{figure*}

\section{Signatures of disturbance}

\label{Signatures of disturbance}

In this section, we extend our analysis to examine several disturbance indices across our galaxy populations and their relationship with stellar mass.  Specifically, we focus on three indices: the residual flux fraction (\emph{RFF}), the asymmetry of the galaxy image ($A_{\mathrm{gal}}$), and the asymmetry of the residual image after model subtraction ($A_{\mathrm{res}}$).  Together, these indices provide complementary insight beyond the basic structural parameters ($R_{\rm e}$ and $n$).  They probe potential structural disturbances by capturing deviations from smooth models, overall asymmetry, and subtle residual features that may indicate recent or ongoing dynamical processes (e.g.\ mergers).  In the following analysis, all measurements are determined at $2\,\mu\rm{m}$ ($F200W$), though broadly consistent results are obtained across all eight PRIMER-UDS \emph{JWST}/NIRCam wavebands ($0.9$--$4.4\,\mu\rm{m}$).

We also verify that using a fixed observed-frame waveband does not affect our conclusions by performing parallel analyses constructed using the PRIMER waveband closest to a rest-frame wavelength of $0.8\rm\,\mu{m}$.  We find no significant differences in the resulting trends for any of the disturbance metrics.  In the following, for brevity, we only report the rest-frame results for $A_{\mathrm{res}}$, which we show in Appendix~\ref{Supplementary Figures and Tables} as a representative case (Fig.~\ref{ares-vs-mass-restframe}).

\subsection{Residual flux fraction (RFF)}

\label{Residual flux fraction}

The residual flux fraction (\emph{RFF}) is a quantitative measure of the amount of flux that remains in an image after subtracting a galaxy model, typically a two-dimensional S\'ersic fit \citep{Blakeslee_etal:2006, Hoyos_etal:2011, Hoyos_etal:2012}.  It represents the fraction of light in the residual image that cannot be explained by background noise fluctuations.  The \emph{RFF} is commonly used to assess structural disturbances in galaxies, such as asymmetric distributions, clumps, or features not captured by the symmetric model.  Following \cite{Hoyos_etal:2012}, we define the \emph{RFF} as
\begin{equation}
\textit{RFF} = \frac{\sum_{\,i,j \in A}\,\left| I_{\,i,j} - I_{\,i,j}^{\,\mathrm{Model}} \right| - 0.8\times\sum_{\,i,j \in A}\,\sigma({\mathrm{\scriptstyle Bkg}})_{\,i,j}}{\sum_{\,i,j \in A}\,I_{\,i,j}^{\,\mathrm{Model}}},
\end{equation}
where $I_{\,i,j}$ represents the original image flux, $I_{\,i,j}^{\rm\, Model}$ is the 2D S\'ersic model created by {\sc{galfitm}}, and $\sigma({\mathrm{\scriptstyle Bkg}})_{\,i,j}$ is the background rms map determined by {\sc{SExtractor}} (see Section~\ref{Source detection and sky subtraction}).  In this work, the summations run over all pixels ($i$,$j$) within an aperture $A$ defined by the {\sc SExtractor} Kron aperture, which typically encloses more than $90$ per cent of the galaxy's total flux.  The factor of $0.8$ in the background correction ensures that the expected value of the \emph{RFF} is zero for an image containing only Gaussian noise of constant variance.  This factor is derived from the integral
\begin{equation}
0.8 = \sqrt{\frac{1}{2\pi}}\int^\infty_{-\infty} \lvert x \rvert \times e^{-x^2/2}\,{\mathrm{d}}x,
\end{equation}
which is the expectation value of the absolute value of a Gaussian random variable \citep[see][for further details]{Hoyos_etal:2012}.

In Fig.~\ref{rff-vs-mass}, we present the \emph{RFF} as a function of stellar mass for star-forming, passive, and PSB galaxies across three redshift intervals: $0.5 < z < 1$, $1 < z < 2$, and $2 < z < 3$.  Here the \emph{RFF} values are measured at $2\,\mu\mathrm{m}$ ($F$200$W$), and we show both individual galaxies and the median \emph{RFF} calculated in stellar mass bins of $0.25$ dex.  For each population and redshift interval, we also indicate the $90$ per cent mass completeness limits, determined at the upper bound of the respective redshift interval (see Table~\ref{mass-comp-limits}).  Uncertainties on the median \emph{RFF} ($1\sigma$) correspond to the standard error on the median value.

Across all redshifts ($0.5 < z < 3$), all galaxy populations exhibit a general increase in \emph{RFF} with increasing stellar mass.  Star-forming galaxies consistently show higher \emph{RFF} values than passive and PSB galaxies of the same stellar mass, reflecting their typically more complex and clumpy sub-structures, such as spiral arms and star-forming regions.  In contrast, passive galaxies tend to have lower \emph{RFF} values, indicative of their smooth stellar distributions.  Notably, PSBs are largely indistinguishable from the passive population in terms of their \emph{RFF}, showing no significant excess residual flux that would imply major structural disturbances or unexplained light.  This is supported by K-S tests, which find no statistically significant difference between the \emph{RFF} distributions of PSBs and passive galaxies across all redshifts and stellar masses ($p > 0.05$).  This suggests that PSBs, despite their recent quenching, do not exhibit significantly enhanced structural irregularities beyond those found in passive galaxies.

Given the sensitivity of the \emph{RFF} to residual structure, particularly in the central regions of galaxies, we assessed the potential impact of PSF mismatches on our measurements.  The careful construction of the PSF in our analysis minimizes this concern; however, as a robustness check, we repeated the analysis with the central $5\times5$ pixel region removed.  This had no significant effect on the results, confirming that our \emph{RFF} measurements are not biased by PSF-related artefacts.

\subsection{Asymmetry ($A_{\rm gal}$ and $A_{\rm res}$)}

\label{Asymmetry}

Asymmetry quantifies the extent to which a galaxy's light distribution deviates from perfect 180-degree rotational symmetry.  It is a widely used, non-parametric morphological indicator that is particularly sensitive to disturbances such as tidal features, irregular star formation, or ongoing mergers.  It is characterized by the asymmetry parameter (denoted \emph{A}), which measures the difference between an image and its 180-degree rotation about the point that minimizes this difference \citep{Bershady_etal:2000, Conselice:2003}.  Mathematically, it is defined as
\begin{equation}
A = \left(\frac{\sum_{\,i,j \in A}\left| I_{\,i,j} - I_{\,i,j}^{\,180}\right|}{\sum_{\,i,j \in A}\left| I_{\,i,j} \right|}\right)_{\mathrm{min}} - \ \ \left(\frac{\sum_{\,i,j \in B}\left| B_{\,i,j} - B_{\,i,j}^{\,180}\right|}{\sum_{\,i,j \in A}\left| I_{\,i,j}\right|}\right)_{\mathrm{min}}
\end{equation}
where $I_{\,i,j}$ is the original image and $I_{\,i,j}^{180}$ is its 180-degree rotation about a centre that minimizes the asymmetry.  The second term corrects for background noise, where $B_{\,i,j}$ is a background region and $B_{\,i,j}^{180}$ its 180-degree rotation.  In this work, as with \emph{RFF}, the summations run over all pixels ($i$,$j$) within an aperture $A$ defined by the {\sc SExtractor} Kron aperture, which typically encloses more than $90$ per cent of the galaxy's total flux.  The background summation is performed over the entire background region $B$.

We consider two forms of asymmetry: the asymmetry in the galaxy ($A_{\rm gal}$) and the asymmetry in the residual image after subtraction of the galaxy's {\sc{galfitm}} model ($A_{\rm res}$).  Following \citet{Hoyos_etal:2012}, the minimization process differs slightly between these measurements.  For $A_{\rm gal}$, the centre of rotation is allowed to vary within $\pm9$ pixels from the {\sc{galfitm}} luminosity-weighted centroid, while for $A_{\rm res}$, the search radius is reduced to $\pm4$ pixels.  In both cases, the background correction term undergoes the same minimization process, with background regions selected as the largest nearby blank sky areas identified using dilated {\sc{SExtractor}} segmentation maps.  The measured asymmetry values for real galaxies ($A_{\rm gal}$) typically range from $0$ to $0.6$, with most galaxies having values around $0.2$.  The asymmetry values for residual images ($A_{\rm res}$) are typically higher, ranging from approximately $0$ to $1.2$, reflecting a heightened sensitivity to structural complexities not captured by the S\'ersic model.

Since the asymmetry metrics $A_{\rm gal}$ and $A_{\rm res}$ are sensitive to residual structures, particularly near the galaxy centres, even a slight PSF mismatch has the potential to bias the measurements.  Although our PSF construction is designed to minimize such effects, we test the robustness of our measurements by evaluating the impact of excluding the central $5\times5$ pixel region.  This exclusion produces no significant change in the measured asymmetries, demonstrating that the measurements are not systematically affected by PSF-related artefacts.

\begin{figure*}
\includegraphics[width=0.85\textwidth]{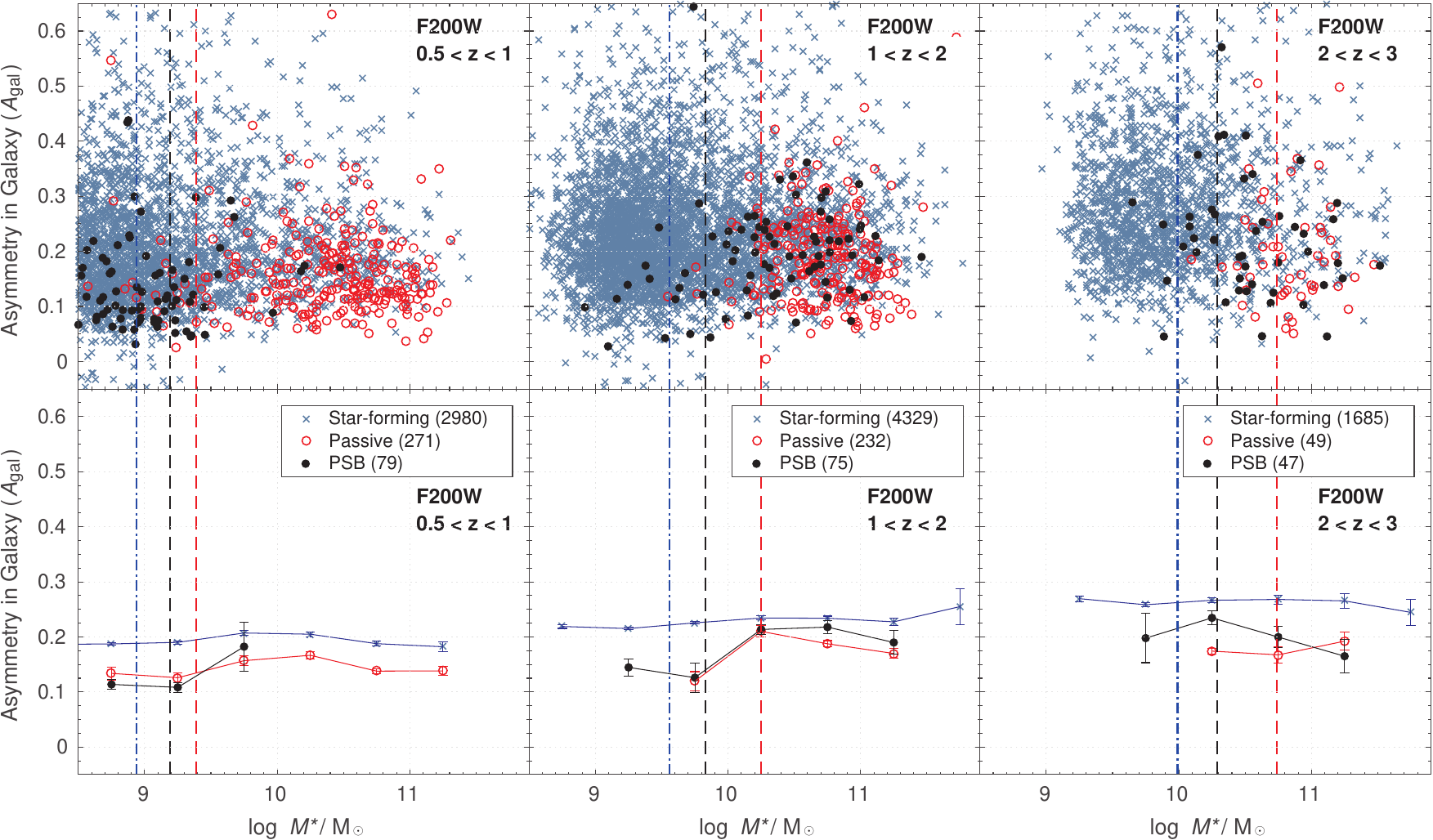}
\centering
\caption{\label{agal-vs-mass} The asymmetry in the galaxy ($A_{\rm gal}$), determined at $2\rm\,\mu{m}$ ($F200W$), as a function of stellar mass for star-forming (blue crosses), passive (red circles) and PSB galaxies (black points).  \emph{Top row}: stellar mass vs.\ $A_{\rm gal}$ for individual galaxies across three redshift intervals: $0.5 < z < 1$ (left-hand panel), $1 < z < 2$ (centre panel) and $2 < z < 3$ (right-hand panel).  \emph{Bottom row}: the corresponding median $A_{\rm gal}$ as a function of stellar mass for each population and redshift interval.  Median $A_{\rm gal}$ values are computed in stellar mass bins of $0.5$~dex, and only shown for bins containing more than two galaxies.  Respective sample sizes are shown in the legend.  The vertical lines represent the 90 per cent mass completeness limits for each population, determined at the upper limit of the respective redshift interval: star-forming (blue dot-dashed), passive (red dashed) and PSB (black dashed).  Errors in the median $A_{\rm gal}$ ($1\sigma$) represent the standard error on the median value.}
\end{figure*}

\begin{figure*}
\includegraphics[width=0.85\textwidth]{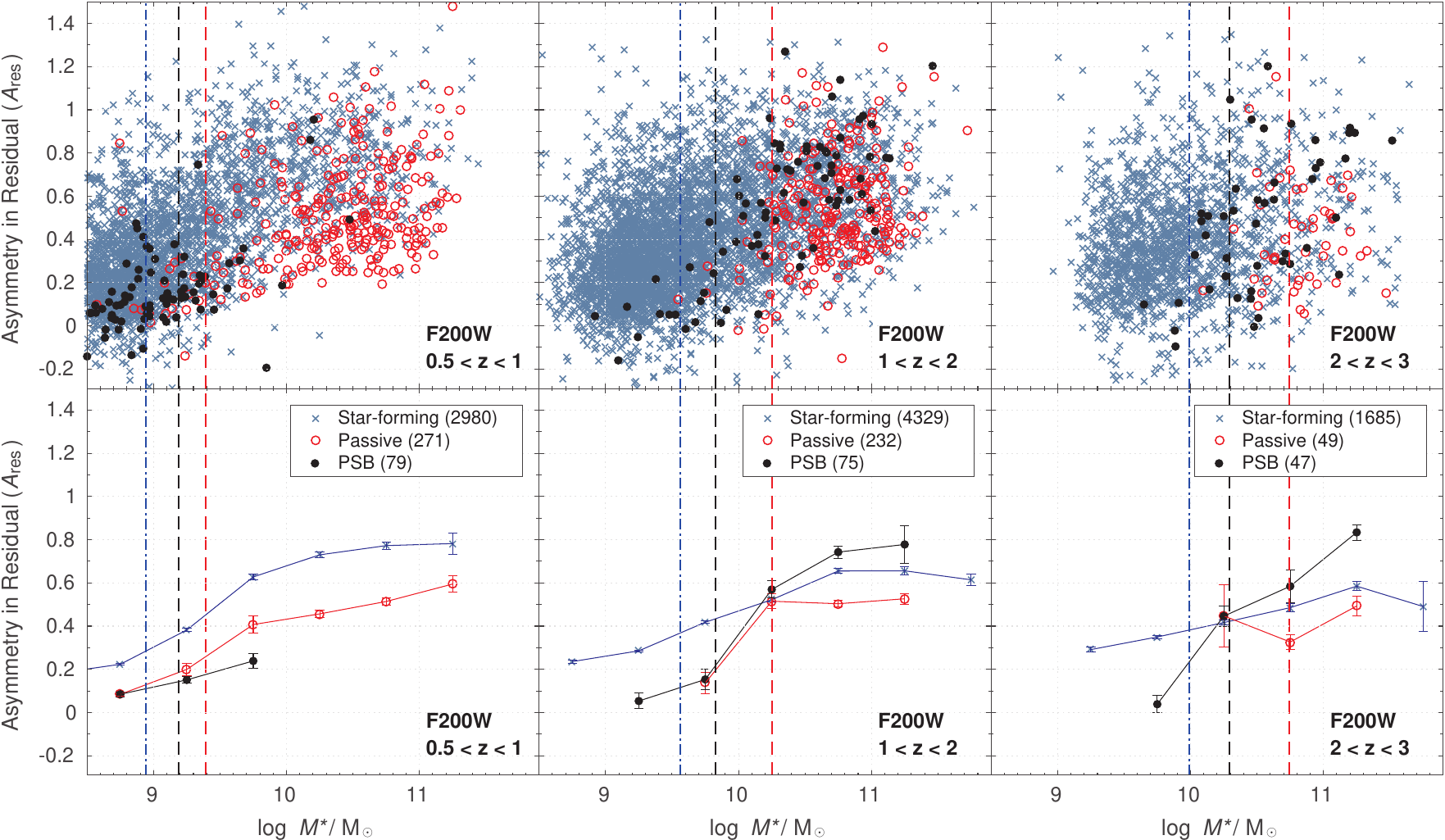}
\centering
\caption{\label{ares-vs-mass} The asymmetry in the residual ($A_{\rm res}$), determined at $2\rm\,\mu{m}$ ($F200W$), as a function of stellar mass for star-forming (blue crosses), passive (red circles) and PSB galaxies (black points).  \emph{Top row}: stellar mass vs.\ $A_{\rm res}$ for individual galaxies across three redshift intervals: $0.5 < z < 1$ (left-hand panel), $1 < z < 2$ (centre panel) and $2 < z < 3$ (right-hand panel).  \emph{Bottom row}: the corresponding median $A_{\rm res}$ as a function of stellar mass for each population and redshift interval.  Median $A_{\rm res}$ values are computed in stellar mass bins of $0.5$~dex, and only shown for bins containing more than two galaxies.  Respective sample sizes are shown in the legend.  The vertical lines represent the 90 per cent mass completeness limits for each population, determined at the upper limit of the respective redshift interval: star-forming (blue dot-dashed), passive (red dashed) and PSB (black dashed).  Errors in the median $A_{\rm res}$ ($1\sigma$) represent the standard error on the median value.  At $z > 1$, we find evidence for an elevated $A_{\rm res}$ in PSB galaxies at $\log M_*/{\rm M_{\odot}} > 10.25$.}
\end{figure*}

Figs.~\ref{agal-vs-mass} and \ref{ares-vs-mass} show our asymmetry measurements, $A_{\rm gal}$ and $A_{\rm res}$, respectively, as a function of stellar mass for various galaxy populations.  We present results for star-forming, passive, and PSB galaxies across three redshift intervals: $0.5 < z < 1$, $1 < z < 2$, and $2 < z < 3$.  In each case, the asymmetry values are measured at $2\,\mu\mathrm{m}$ ($F$200$W$), and we show both individual galaxies and the median asymmetry calculated in stellar mass bins of $0.25$ dex.  For each population and redshift interval, we also indicate the $90$ per cent mass completeness limits, determined at the upper bound of the respective redshift interval (see Table~\ref{mass-comp-limits}).  Uncertainties on the median asymmetry ($1\sigma$) correspond to the standard error on the median value.

For galaxy asymmetry ($A_{\rm gal}$), across all redshifts ($0.5 < z < 3$) we find no significant dependence on stellar mass for any galaxy population (see Fig.~\ref{agal-vs-mass}).  In contrast, there is a clear dependence on galaxy type: at fixed stellar mass, star-forming galaxies consistently exhibit higher $A_{\rm gal}$ values than quenched systems, consistent with established results \cite[e.g.][]{Conselice_etal:2000}.  Although this trend may reflect intrinsic structural differences, at these redshifts it could also be influenced by observational bias.  Star-forming galaxies tend to have more extended light profiles, which can make asymmetric features easier to detect.  At high redshift, galaxies subtend relatively small angular sizes, allowing the PSF to efficiently smooth substructure.  This effect can suppress measured asymmetry in compact systems, such as passive and PSB galaxies, while leaving asymmetries more detectable in the more extended star-forming population.  However, star-forming galaxies also show a clear redshift evolution, with markedly higher $A_{\rm gal}$ values at earlier epochs.  This trend aligns with previous results \cite[e.g.][]{Whitney_etal:2021, Dolfi_etal:2025} and is highly significant ($p < 10^{-7}$ from K-S tests).  Therefore, since high-redshift star-forming galaxies are typically smaller and hence more susceptible to PSF smoothing, the presence of stronger asymmetries at early times suggests that their elevated $A_{\rm gal}$ values are predominantly intrinsic rather than observational in origin.

With respect to quenched systems, both passive and PSB galaxies present low $A_{\rm gal}$ values across all redshifts ($0.5 < z < 3$), and with no significant redshift evolution.  Moreover, at all epochs, PSBs are largely indistinguishable from the passive population in terms of their $A_{\rm gal}$ values, showing no significant excess that would imply major structural disturbances indicative of ongoing dynamical processes.  This conclusion is supported by K–S tests, which reveal no statistically significant differences between the $A_{\rm gal}$ distributions of PSBs and passive galaxies at any redshift or stellar mass ($p > 0.05$).

\begin{figure*}
\includegraphics[width=0.99\textwidth]{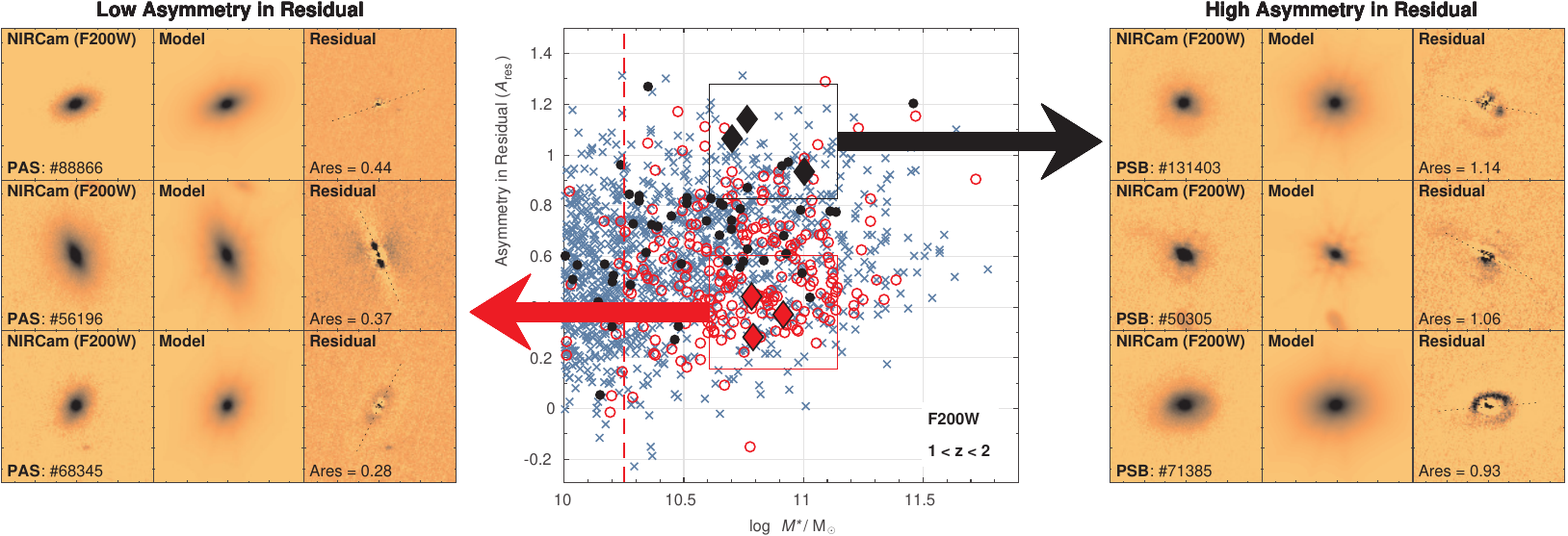}
\centering
\vspace{-0.1cm}
\caption{\label{ares_exmaples} Typical examples of the structural features underlying the observed trends in residual asymmetry.  \emph{Centre panel}: the residual asymmetry as a function of stellar mass for $1 < z < 2$ and $M_* > 10^{10}\,\rm{M_{\odot}}$ (as shown in Fig.~\ref{ares-vs-mass}, top-centre panel).  \emph{Left panels}: three example passive galaxies with low residual asymmetry. Each row displays the $F200W$ science image, the corresponding S\'ersic model, and the resulting residual.  Although these objects exhibit significant residual flux, the structure is clearly axisymmetric.  \emph{Right panels}: three example PSBs with high residual asymmetry, again showing the $F200W$ image, S\'ersic model, and residual.  In these cases, the residual light is visibly asymmetric, consistent with late-stage dynamical processes.  All images measure $3.6\arcsec\times3.6\arcsec$.}
\end{figure*}

\begin{figure}
\includegraphics[width=0.475\textwidth]{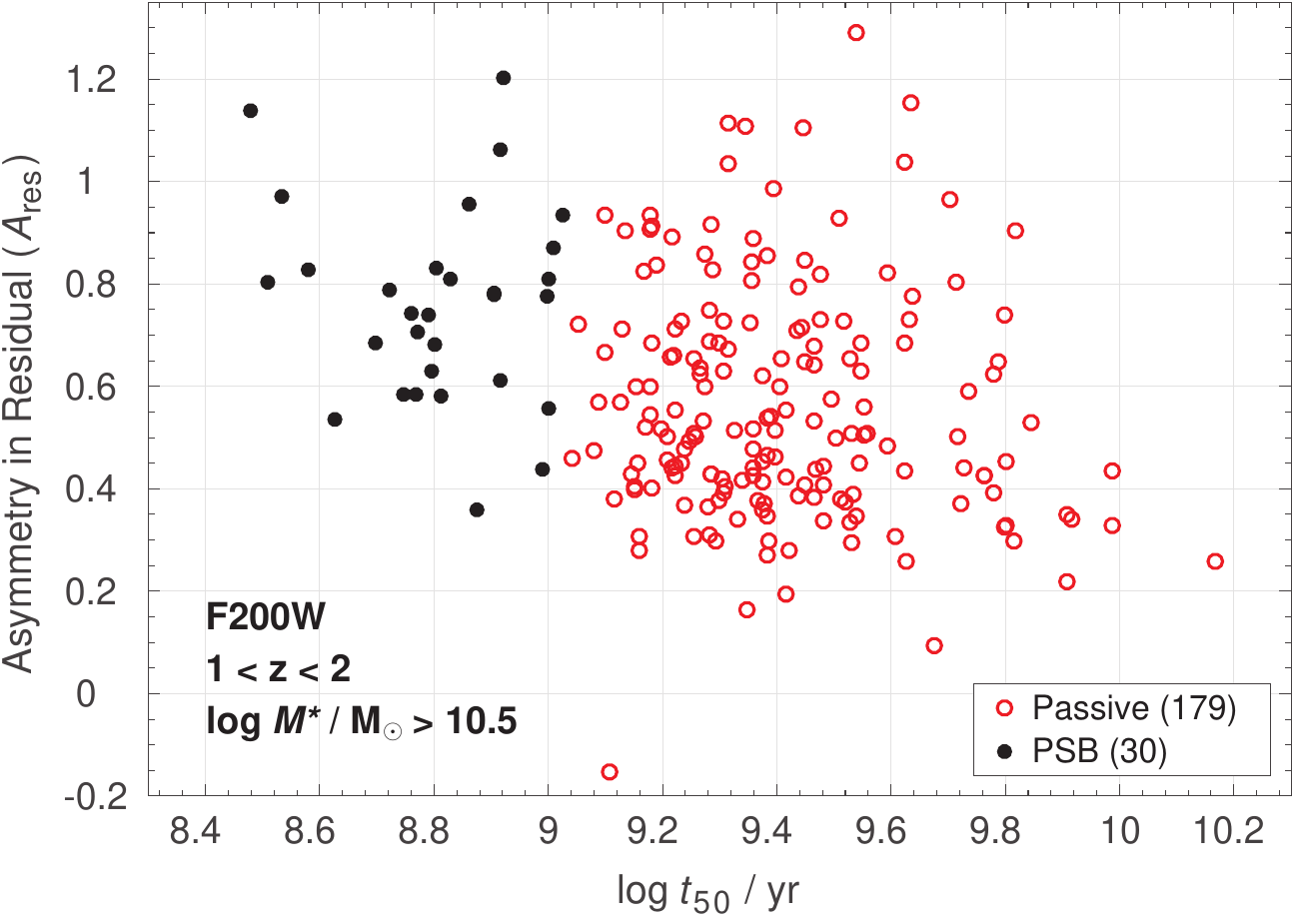}
\centering
\caption{\label{ares-vs-time} The asymmetry in the residual galaxy light ($A_{\rm res}$), determined at $2\rm\,\mu{m}$ ($F200W$), as a function of the median stellar age ($t_{50}$), as defined by 
\citet[]{Belli_etal:2019}. Passive galaxies and PSBs are shown, selected over the redshift range $1<z<2$, with stellar masses $\log M_*/{\rm M_{\odot}} > 10.5$.  We find evidence for a decline of asymmetry with burst age, with a correlation coefficient  $\rho = -0.292$ (significance $p < 10^{-4}$).  We caution, however, that the ages are derived from photometric data, so may be affected by degeneracies between the age and strength of the burst.}
\end{figure}

To probe more subtle structural irregularities, we turn to residual asymmetry ($A_{\rm res}$), which is sensitive to more subtle departures from a smooth light distribution (see Fig.~\ref{ares-vs-mass}).  In contrast to galaxy asymmetry ($A_{\rm gal}$), all galaxy populations show a general increase in residual asymmetry ($A_{\rm res}$) with increasing stellar mass across the full redshift range ($0.5 < z < 3$).  At $0.5 < z < 1$, star-forming galaxies exhibit distinctly higher residual asymmetries than passive and PSB systems, consistent with their clumpier and more irregular substructures, while PSBs at this epoch show comparable, though marginally lower, $A_{\rm res}$ values than passive galaxies at fixed stellar mass.  At $1 < z < 2$, population differences become more pronounced, with PSBs displaying a clear enhancement in $A_{\rm res}$ at the highest stellar masses ($M_* > 10^{10.25}\,\mathrm{M_{\odot}}$), exceeding that of both the passive and star-forming populations.  This result is supported by K-S tests ($p < 5\times10^{-6}$ and $p < 5\times10^{-3}$ for the passive and star-forming comparisons, respectively).  At $z > 2$, we again find massive PSBs exhibit elevated $A_{\rm res}$ relative to the passive population, consistent with the trends at $1 < z < 2$.  However, the statistical significance is weaker ($p < 0.02$), likely due to limited number statistics.  This finding suggests that, although massive PSBs lack prominent structural disturbances, they may host subtle asymmetric components hidden beneath a dominant smooth stellar distribution, potentially indicating late-stage signatures of a major merger.  These results highlight the utility of $A_{\rm res}$ in revealing structural signatures that are otherwise hidden in the total light distribution.

We verify that these fiducial $F200W$ $A_{\rm{res}}$ results are robust to rest-frame wavelength dependence by constructing relations using the PRIMER waveband closest to a rest-frame wavelength of $0.8\,\rm\mu{m}$. We find that this yields consistent results, confirming that the resulting trends are not significantly affected by variation in the effective rest-frame wavelength across the sample (see Appendix~\ref{Supplementary Figures and Tables}; Fig.~\ref{ares-vs-mass-restframe}).

We note that asymmetry measurements have been shown to depend on imaging depth and resolution \citep[e.g.][]{Sazonova_etal:2025}.  To assess whether the elevated $A_{\rm res}$ values observed in PSBs could be driven by differences in data quality, we examine both the aperture S/N and the median per-pixel S/N within fixed $1$ arcsec apertures as a function of stellar mass and redshift.  The aperture S/N provides a measure of the overall depth of the data, while the per-pixel S/N more directly reflects the noise properties relevant for pixel-based asymmetry measurements.  We find that PSB and passive galaxies exhibit consistent S/N distributions across all stellar masses and redshifts, including in the regime where the $A_{\rm res}$ enhancement is observed ($\log M_*/{\rm M_{\odot}} > 10.25$, $1 < z < 2$), with the distributions of both aperture S/N and per-pixel S/N being statistically indistinguishable based on K-S tests ($p > 0.05$), a result that remains unchanged when adopting a more restrictive mass cut of $\log M_*/{\rm M_{\odot}} > 10.6$.  Taken together, these results indicate that the enhanced residual asymmetry in PSBs is not driven by differences in the S/N between populations.  We note that resolution can also affect asymmetry measurements, generally suppressing values in more compact systems.  However, this is not expected to be a significant issue in our analysis, as the vast majority of galaxies in our sample are well resolved relative to the PSF.  Furthermore, PSBs -- the most compact population in our sample -- exhibit enhanced asymmetry, which argues against resolution effects being responsible for the observed trends.

To illustrate the structural features underlying the residual asymmetry trends at $z > 1$, in Fig.~\ref{ares_exmaples} we present three representative PSBs with high $A_{\rm res}$ values together with three passive galaxies exhibiting low $A_{\rm res}$ values.  For each system, we show the $F200W$ science image, the best-fitting S\'ersic model, and the residual image from which the residual asymmetry is computed.  In the PSBs with high $A_{\rm res}$, the residual maps reveal clear asymmetric substructures (e.g.\ faint tidal features, off-centre light concentrations, or irregular low-surface-brightness components) that are largely suppressed when considering the total galaxy light (i.e.\ measurements of $A_{\rm gal}$; see Fig.~\ref{agal-vs-mass}).  In contrast, although the passive galaxies may also show substantial residual structure, it is distinctly axisymmetric and reflects recognizable structural components not captured by a single S\'ersic model (e.g.\ stellar bars and ansae), resulting in low $A_{\rm res}$ despite their comparable stellar masses. These visual examples demonstrate how $A_{\rm res}$ isolates subtle morphological signatures that are not evident in the science images alone, and help contextualize the elevated asymmetry observed in massive PSBs at $z > 1$.

\subsection{Asymmetry as a function of median stellar age}

\label{Asymmetry_time}

The above analysis suggests that massive PSBs show significantly higher values for residual asymmetry ($A_{\rm res}$) compared to older passive galaxies.  This trend may point to a merger-driven origin, consistent with \cite{Ellison_etal:2022}, who found a strong association between rapidly-quenched galaxies and post-merger signatures in the nearby Universe.  Simulations of high-redshift galaxies suggest that merger features are detectable for up to $\sim1$~Gyr in rest-frame optical observations \citep{Whitney_etal:2021}.  If post-merger quenching occurs within $\sim0.5$~Gyr of coalescence \citep{Ellison_etal:2024}, we might therefore expect the most recently quenched galaxies to display the highest incidence of merger signatures.

In Fig.~\ref{ares-vs-time}, we present the residual asymmetry ($A_{\rm res}$) as a function of the median stellar age for massive ($M_* > 10^{10.5}\,{\rm M_{\odot}}$) PSB and passive galaxies in the redshift range $1 < z < 2$.  The median stellar age ($t_{50}$) is estimated from the position in the rest-frame  $\mathit{UVJ}$ diagram, using the method outlined in \cite{Belli_etal:2019}.  We find  evidence for a decline in the average asymmetry with median stellar age. Combining PSB and passive galaxies, a Spearman's rank correlation test shows evidence for a negative monotonic decline of asymmetry with $t_{50}$, with a correlation coefficient $\rho = -0.292$, and a low probability of a chance correlation ($p < 10^{-4}$).
To test the robustness of this result to uncertainties in the derived ages, we also perform a bootstrap analysis in which $\log(t_{50})$ values are perturbed within their typical $1\sigma$ uncertainties ($\sim0.13$ dex; \citealt{Belli_etal:2019}), and the correlation is recomputed over $1000$ realisations.  This yields a consistent negative correlation, with a 68 per cent confidence interval of $[-0.327, -0.198]$, indicating that the negative correlation is robust to the adopted age uncertainties.

Though not shown for brevity,  we obtain consistent results for the quenched galaxies at higher redshift ($2<z<3$), which also show evidence for an anticorrelation of asymmetry with median stellar age ($\rho = -0.407$, $p<10^{-3}$). The 68 per cent confidence interval from the bootstrap analysis is $[-0.487, -0.262]$.  We note that the ages obtained from photometric data are also likely to be uncertain, due to degeneracies between the age and the strength of the burst \citep[see][]{Wild_etal:2020}.  A future analysis combining spectroscopic and photometric data would yield more precise star formation histories, which may be possible once larger spectroscopic samples of PSBs are obtained, (e.g.\ from the forthcoming VLT MOONRISE survey; \citealt{Maiolino_etal:2020}).  Such data may allow a more detailed understanding of the structural transformation of galaxies as a function of time during the quenching phase.

\section{Discussion and Conclusions}

\label{Conclusions}

We have undertaken a detailed study of the multiwavelength structure of recently quenched PSBs at $0.5 < z < 3$, combining deep eight-band \emph{JWST}/NIRCam imaging from the PRIMER programme with a robust photometrically selected sample from the UDS field.  Structural parameters were measured from single S\'ersic fits independently across the rest-frame optical–to–near-infrared, and disturbance indices (e.g.\ $\textit{RFF}$ and asymmetry) were systematically quantified.  By comparing these measurements to a large control sample of passive and star-forming galaxies, we have mapped the internal structure, morphological evolution, and subtle signs of dynamical disturbance in PSBs across $0.5 < z < 3$ with unprecedented detail.

Across all redshifts, PSBs exhibit remarkably uniform structural parameters with wavelength.  Both the effective radius ($R_{\rm e}$) and S\'ersic index ($n$) show minimal dependence on rest-frame wavelength, indicating that these systems lack strong stellar population gradients or significant internal dust obscuration.  This uniformity suggests that PSBs are already well mixed and dynamically settled by the time star formation ceases, reinforcing the view that they represent a post-transformation phase in which both quenching and major structural changes are largely complete.

Placed in the broader evolutionary context, we confirm that PSBs follow the established redshift--mass trends reported in previous works \citep[e.g.][]{Whitaker_etal:2012a, Almaini_etal:2017, Maltby_etal:2018, Zhang_etal:2024}.  At $z > 1$, massive PSBs ($M_* > 10^{10}\,{\rm M}_\odot$) are extremely compact spheroids, morphologically similar to the passive population but systematically smaller in size.  Their compactness and high S\'ersic indices point to a formation via rapid, dissipative processes, such as gas-rich major mergers or protogalactic collapse, which simultaneously build dense stellar cores and trigger abrupt quenching by either AGN or starburst driven feedback.  By contrast, at lower redshifts ($0.5 < z < 1$), PSBs are predominantly lower-mass ($M_* < 10^{10}\,{\rm M}_\odot$) systems with disc-dominated morphologies and more moderate S\'ersic indices.  Their structural similarity to low-mass passive discs suggests a gentler, more extended transformation, potentially driven by minor mergers or environmentally induced gas stripping rather than major dynamical upheaval.  Taken together, these results reinforce the growing consensus that there are at least two distinct quenching pathways, which dominate at different epochs and mass regimes.

A novel aspect of our study is the systematic quantification of disturbance indicators, which leads to our key finding that high-mass PSBs at $z > 1$ present a previously unrecognised level of structural disturbance hidden beneath a smooth stellar distribution.  Across all redshifts, we find that, on average, PSBs are morphologically smooth, exhibiting \textit{RFF} values and asymmetries comparable to those of passive galaxies. This smoothness indicates that, for most systems, any strong morphological disturbances have faded by the PSB phase, implying a short relaxation timescale following quenching.  However, we identify a clear enhancement in residual asymmetry among massive ($M_* > 10^{10.25}\,{\rm M}_\odot$) PSBs at $z > 1$.  This subtle but significant signal of structural disturbance suggests that some high-redshift PSBs retain the imprints of late-stage dynamical processes (e.g.\ tidal interactions, or remnant asymmetries from recent major events).  The coexistence of compact, largely relaxed morphologies with faint residual asymmetries supports a picture in which structural transformation is nearly complete, yet the final dynamical settling of stellar mass is still ongoing.  This idea is reinforced by our observation that, when considering the quenched population as a whole (passive + PSB), there is a clear evidence for a  decline in residual asymmetry with increasing median stellar age.

Taken together, these findings point toward two dominant evolutionary pathways for rapid quenching.  At early epochs ($z > 1$), massive PSBs likely originate from highly dissipative and disruptive events that produce compact spheroids and rapidly terminate star formation.  At later times ($z < 1$), lower-mass PSBs appear to form through more gradual, non-violent mechanisms that preserve disc structure and lead to passive, disc-dominated remnants, generally associated with rotational support.  This dual-channel picture is consistent with the evolving physical conditions of the Universe \citep[e.g. declining gas fractions, merger rates, and specific star formation rates; see e.g.][]{Conselice:2014, Madau&Dickinson:2014, Tacconi_etal:2020} suggesting that the mechanisms responsible for rapid quenching evolve alongside the broader galaxy population.

The results presented here highlight the power of deep, high-resolution \textit{JWST} imaging to dissect the internal structure of transitional galaxy populations.  By combining rest-frame optical and near-infrared structural diagnostics with sensitive morphological disturbance measures, we can directly link observed morphology to quenching mechanisms and evolutionary stage.  Future work, incorporating spectroscopic confirmation, stellar kinematics, and spatially resolved color or emission-line mapping, will provide critical constraints on the timescales and physical drivers of the PSB phase.  Together, such analyses will allow a more complete understanding of how galaxies transition from star-forming to quiescent, and of the diverse physical routes through which this transformation proceeds across cosmic time.

\section*{Acknowledgements}

DTM and OA acknowledge the support from STFC grant ST/X006581/1. VW acknowledges the support of the Science and Technologies Facilities Council (ST/Y00275X/1) and Leverhulme Research Fellowship (RF-2024-589/4). JSD acknowledges the support of the Royal Society through the award of a Royal Society Research Professorship.  ET and ACC acknowledge support from a UKRI Frontier Research Guarantee Grant (PI Carnall; grant reference EP/Y037065/1).  We gratefully acknowledge support from the NASA Astrophysics Data Analysis Program (ADAP) under grant 80NSSC23K0495.

\section*{Data Availability}

The data forming the basis of this work is available from public archives, including the Mikulski Archive for Space Telescopes (https://mast.stsci.edu/), and the UKIDSS UDS web page (https://www.nottingham.ac.uk/astronomy/UDS/). Data can also be obtained on reasonable request to the corresponding author.

\bibliographystyle{mnras} \bibliography{DTM_bibtex}

\appendix
\section{Supplementary Figures and Tables}

\label{Supplementary Figures and Tables}

This appendix presents supplementary figures and tables that are referenced in the discussion of the main paper.

\begin{figure*}
\includegraphics[width=0.95\textwidth]{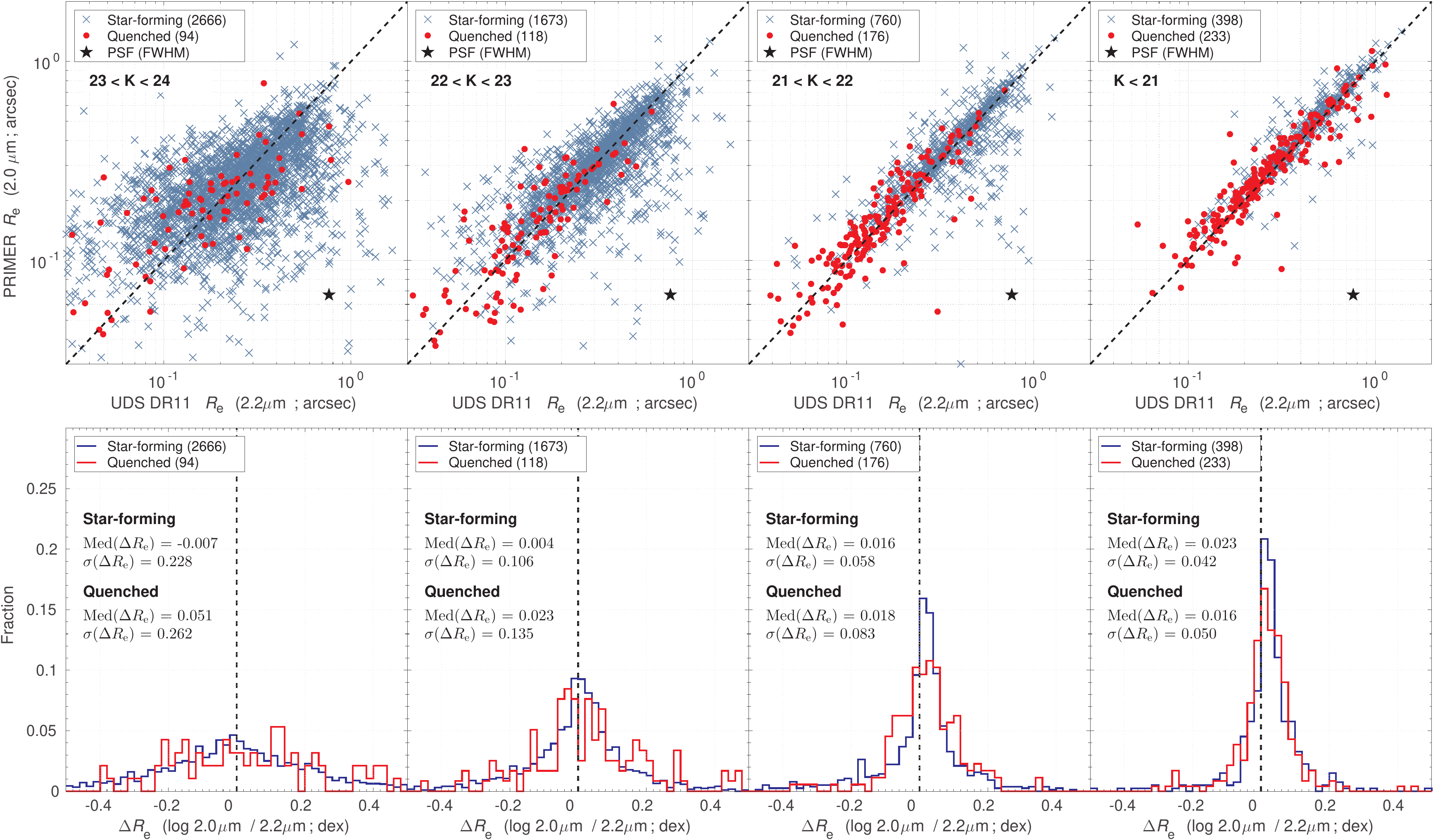}
\centering
\vspace{-0.0cm}
\caption{\label{UDS-re-comp} A comparison of galaxy size ($R_{\rm e}$) measurements from PRIMER-UDS \emph{JWST}/NIRCam $F200W$ imaging ($2.0\rm\,\mu$m; this work) and UDS DR11 UKIRT $K$-band imaging ($2.2\rm\,\mu$m; \citealt{Almaini_etal:2017}).  Comparisons are shown for both star-forming and quenched galaxy populations, where \emph{quenched} refers to the combined passive and PSB populations.  \emph{Top row}: galaxy size ($R_{\rm e}$) comparisons across different $K$-band magnitude ranges.  The black dashed line indicates the one-to-one relation and the PSF FWHM is also shown for reference (black star).  \emph{Bottom row}: the difference ($\Delta R_{\rm e}$) between the PRIMER $2.0\rm\,\mu$m and UDS DR11 $2.2\,\mu$m size measurements across different $K$-band magnitude ranges, where $\Delta R_{\rm e} = {\rm log}\, [R_{\rm e}({\rm PRIMER}) / R_{\rm e}({\rm UDS\ DR11})]$.  Respective sample sizes are shown in the legends.  In each case, the median and scatter in $\Delta R_{\rm e}$ -- Med($\Delta R_{\rm e}$) and $\sigma(\Delta R_{\rm e}$), respectively -- demonstrate there is very good agreement between the PRIMER and UDS DR11 sizes for both star-forming and quenched populations, at least out to $K < 23$.}
\end{figure*}

\begin{figure*}
\includegraphics[width=0.85\textwidth]{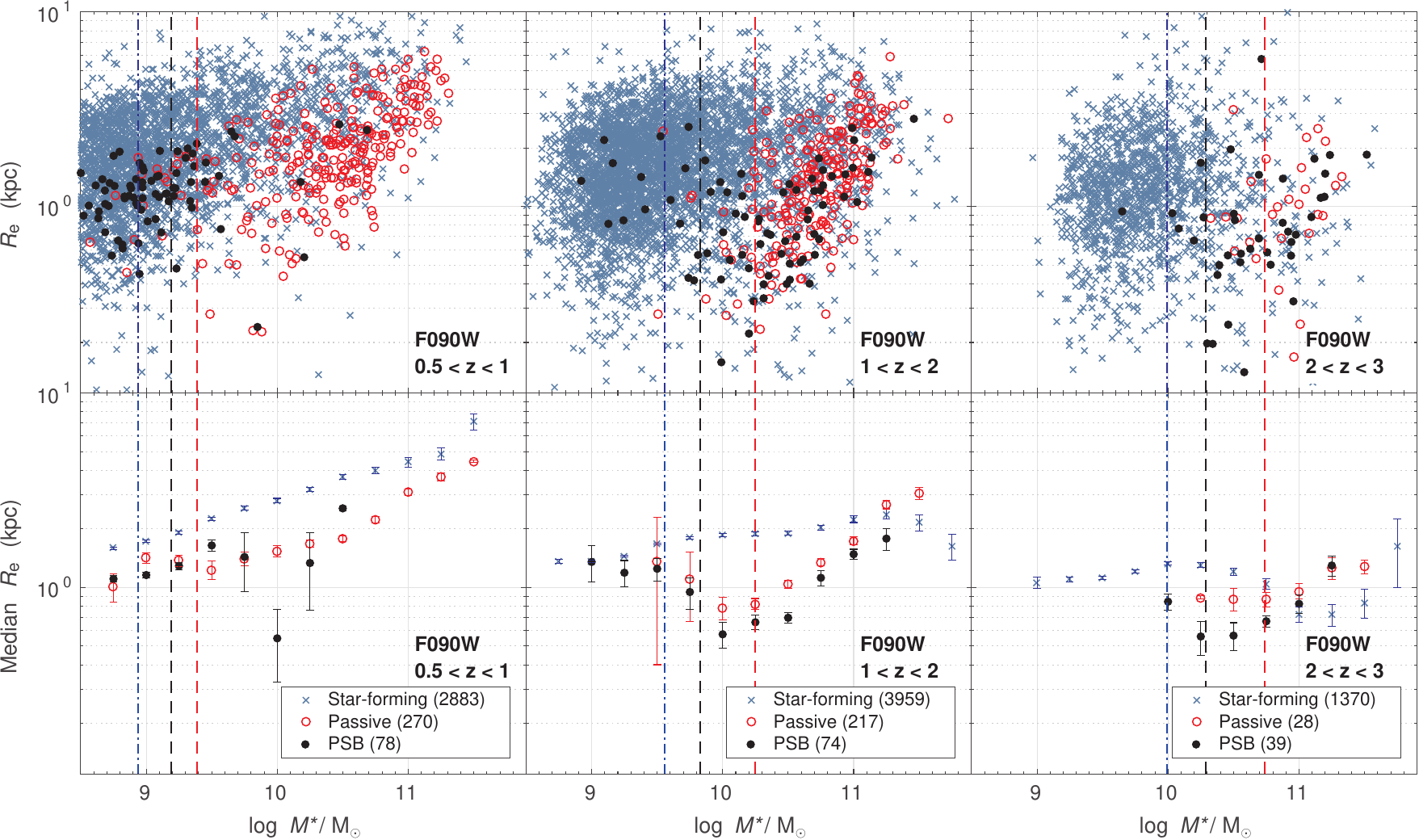}
\centering
\vspace{-0.1cm}
\caption{\label{mass-size-f090w} The evolution of the stellar-mass--size relation for different populations using sizes ($R_{\rm e}$) determined at $0.9\rm\,\mu$m ($F090W$).  \emph{Top~row}: stellar mass vs.\ $R_{\rm e}$ for individual galaxies across three redshift intervals: $0.5 < z < 1$ (left-hand panel), $1 < z < 2$ (centre panel) and $2 < z < 3$ (right-hand panel).  \emph{Bottom row}: the corresponding median $R_{\rm e}$ as a function of stellar mass for each population and redshift interval.  Median $R_{\rm e}$ values are computed in stellar mass bins of $0.25$~dex, and only shown for bins containing more than two galaxies.  Respective sample sizes are shown in the legend.  The vertical lines represent the 90 per cent mass completeness limits for each population, determined at the upper limit of the respective redshift interval (see Table~\ref{mass-comp-limits}): star-forming (blue dot-dashed), passive (red dashed) and PSB (black dashed).  Errors in the median $R_{\rm e}$ ($1\sigma$) represent the standard error on the median value.}
\end{figure*}

\begin{figure*}
\includegraphics[width=0.85\textwidth]{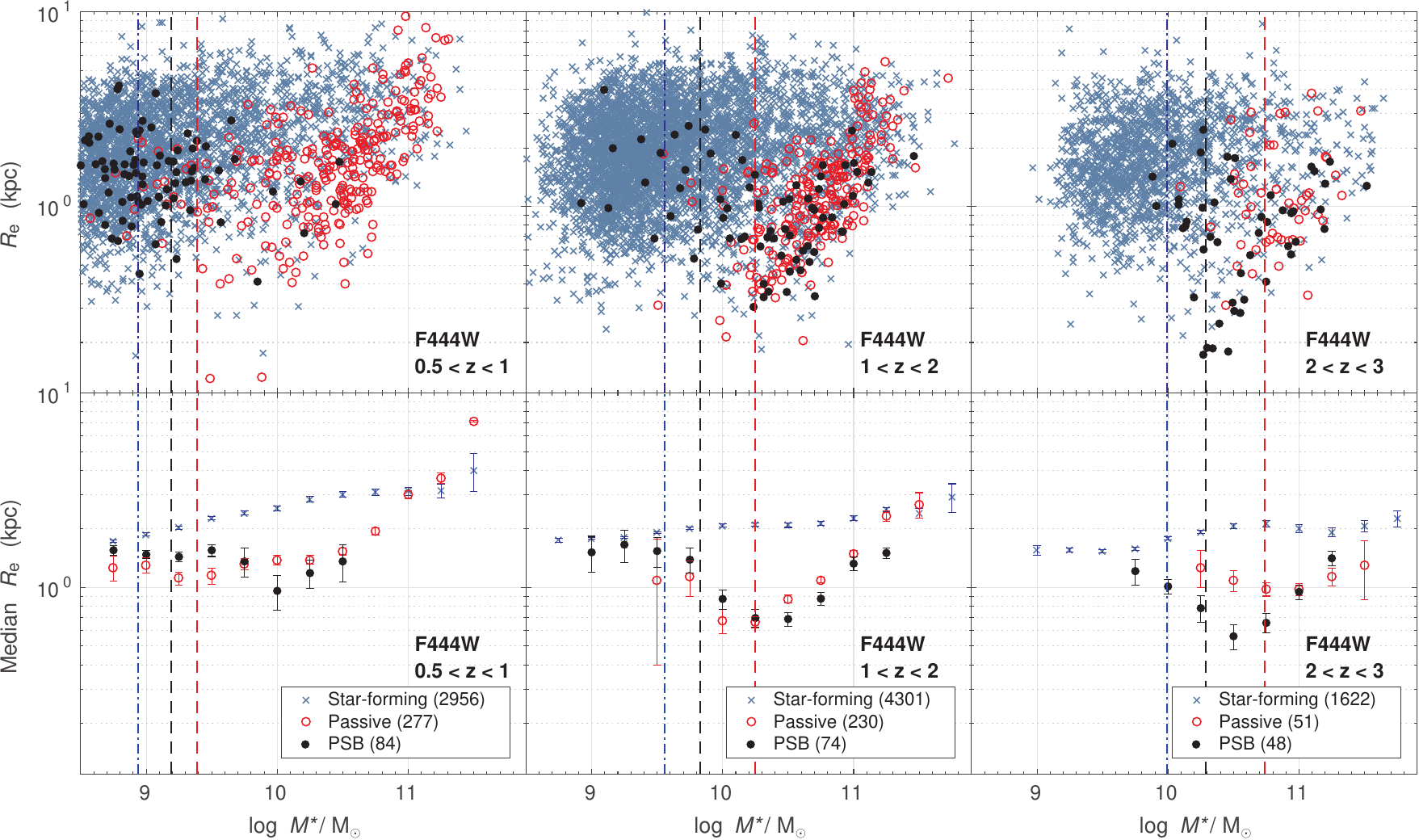}
\centering
\vspace{-0.1cm}
\caption{\label{mass-size-f444w} The evolution of the stellar-mass--size relation for different populations using sizes ($R_{\rm e}$) determined at $4.4\rm\,\mu$m ($F444W$).  \emph{Top~row}: stellar mass vs.\ $R_{\rm e}$ for individual galaxies across three redshift intervals: $0.5 < z < 1$ (left-hand panel), $1 < z < 2$ (centre panel) and $2 < z < 3$ (right-hand panel).  \emph{Bottom row}: the corresponding median $R_{\rm e}$ as a function of stellar mass for each population and redshift interval.  Median $R_{\rm e}$ values are computed in stellar mass bins of $0.25$~dex, and only shown for bins containing more than two galaxies.  Respective sample sizes are shown in the legend.  The vertical lines represent the 90 per cent mass completeness limits for each population, determined at the upper limit of the respective redshift interval (see Table~\ref{mass-comp-limits}): star-forming (blue dot-dashed), passive (red dashed) and PSB (black dashed).  Errors in the median $R_{\rm e}$ ($1\sigma$) represent the standard error on the median value.}
\end{figure*}

\begin{figure*}
\includegraphics[width=0.90\textwidth]{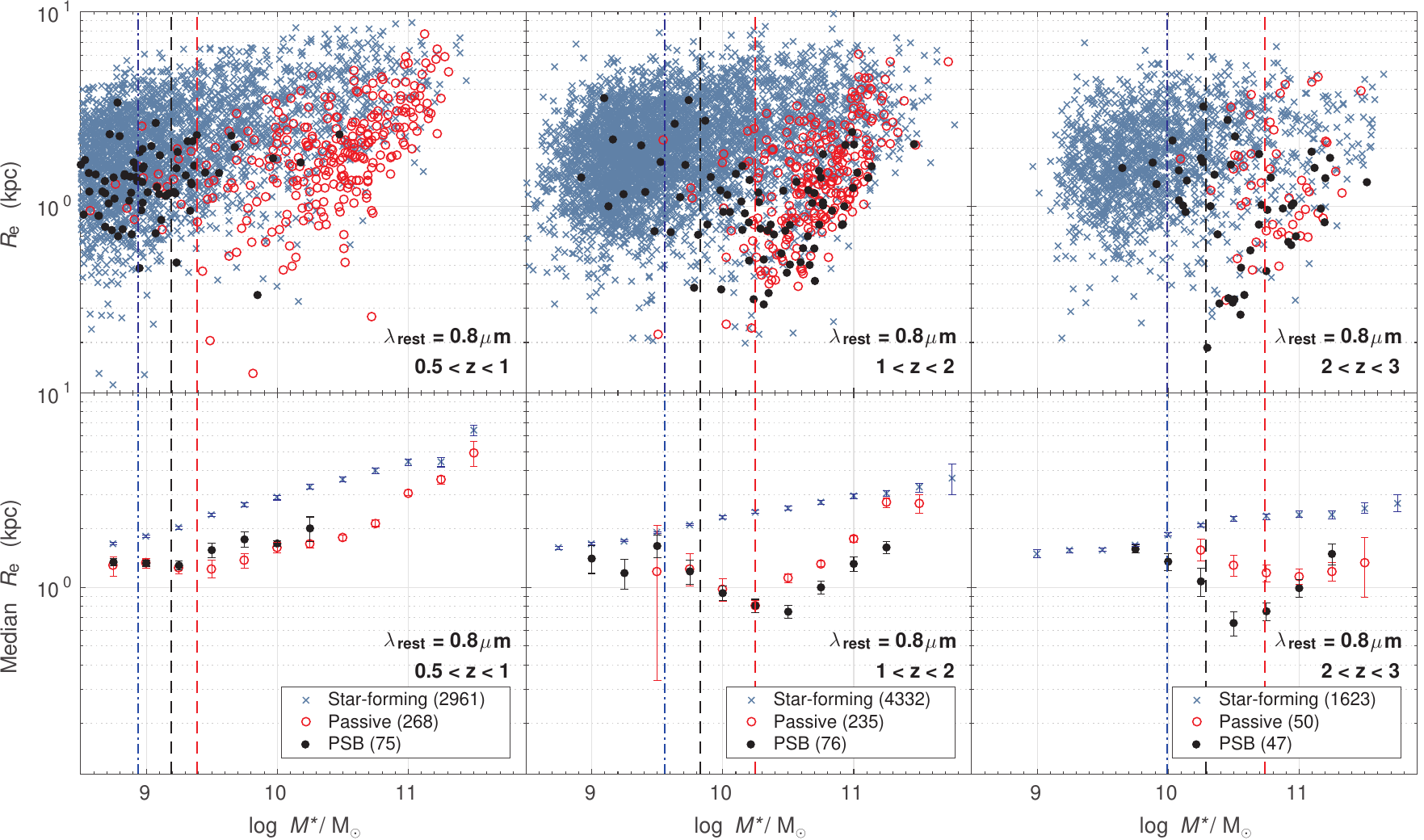}
\centering
\vspace{-0.1cm}
\caption{\label{mass-size-restframe} The evolution of the stellar-mass--size relation for different populations using sizes ($R_{\rm e}$) determined using the PRIMER waveband closest to a rest-frame wavelength of $0.8\,\mu\mathrm{m}$.  The layout, symbols, and binning are identical to Fig.~\ref{mass-size-f200w}, which uses observed-frame $F200W$ measurements.  The resulting relations are fully consistent with those derived only from the $F200W$ imaging, demonstrating that the inferred trends are not driven by wavelength-dependent effects.}
\end{figure*}

\begin{table*}
\centering
\begin{minipage}{173mm}
\centering
\caption{\label{struct-vs-wave-table} Typical structural properties of various galaxy populations measured at different wavelengths ($\lambda_{\rm obs} = 0.9$--$4.4\,\mu\mathrm{m}$).  The median structural parameters ($R_{\rm e}$, $n$) are shown for star-forming, passive, and PSB galaxies across three redshift intervals: $0.5 < z < 1$, $1 < z < 2$, and $2 < z < 3$.  Errors represent the $1\sigma$ uncertainty in the median, as determined from $100$ bootstrapped simulations.}
\begin{tabular}{lccccccccc}
\hline
{Redshift} &{Structural} &\multicolumn{8}{c}{Wavelength (NIRCam filter)} \\[1ex]
{Interval} & {Property} &{$0.9\,\mu\mathrm{m}$} &{$1.2\,\mu\mathrm{m}$} &{$1.5\,\mu\mathrm{m}$} &{$2.0\,\mu\mathrm{m}$} &{$2.8\,\mu\mathrm{m}$} &{$3.6\,\mu\mathrm{m}$} &{$4.1\,\mu\mathrm{m}$} &{$4.4\,\mu\mathrm{m}$}\\
{} & {} &{($F$090$W$)} &{($F$115$W$)} &{($F$150$W$)} &{($F$200$W$)} &{($F$277$W$)} &{($F$356$W$)} &{($F$410$M$)} &{($F$444$W$)} \\
\hline \\[-1.5ex]
\multicolumn{10}{c}{\rule[0.5ex]{0.3\linewidth}{0.5pt}\quad{Star-forming}\quad\rule[0.5ex]{0.3\linewidth}{0.5pt}} \\[1ex]
{$0.5 < z < 1$} &{$R_{\rm e}$ (kpc)} &{$2.69\pm0.06$} &{$2.76\pm0.06$} &{$2.76\pm0.06$} &{$2.64\pm0.05$} &{$2.55\pm0.05$} &{$2.50\pm0.05$} &{$2.45\pm0.05$} &{$2.50\pm0.05$} \\
{}              &{$n$}               &{$1.26\pm0.03$} &{$1.29\pm0.03$} &{$1.35\pm0.02$} &{$1.41\pm0.02$} &{$1.35\pm0.02$} &{$1.36\pm0.02$} &{$1.40\pm0.02$} &{$1.41\pm0.02$} \\[0.5ex]
{$1 < z < 2$}   &{$R_{\rm e}$ (kpc)} &{$1.94\pm0.04$} &{$2.31\pm0.04$} &{$2.62\pm0.05$} &{$2.53\pm0.05$} &{$2.34\pm0.04$} &{$2.20\pm0.04$} &{$2.12\pm0.04$} &{$2.13\pm0.04$} \\
{}              &{$n$}               &{$1.47\pm0.04$} &{$1.44\pm0.04$} &{$1.46\pm0.03$} &{$1.48\pm0.03$} &{$1.46\pm0.03$} &{$1.48\pm0.03$} &{$1.51\pm0.03$} &{$1.54\pm0.03$} \\[0.5ex]
{$2 < z < 3$}   &{$R_{\rm e}$ (kpc)} &{$1.08\pm0.07$} &{$1.52\pm0.08$} &{$2.02\pm0.09$} &{$2.39\pm0.08$} &{$2.35\pm0.08$} &{$2.20\pm0.08$} &{$2.11\pm0.08$} &{$2.07\pm0.08$} \\
{}              &{$n$}               &{$1.52\pm0.09$} &{$1.55\pm0.08$} &{$1.65\pm0.07$} &{$1.34\pm0.06$} &{$1.32\pm0.06$} &{$1.35\pm0.05$} &{$1.42\pm0.06$} &{$1.48\pm0.06$} \\[1ex]
\multicolumn{10}{c}{\rule[0.5ex]{0.3\linewidth}{0.5pt}\quad{Passive}\quad\rule[0.5ex]{0.3\linewidth}{0.5pt}} \\[1ex]
{$0.5 < z < 1$} &{$R_{\rm e}$ (kpc)} &{$1.97\pm0.08$} &{$2.00\pm0.08$} &{$1.90\pm0.08$} &{$1.78\pm0.07$} &{$1.73\pm0.07$} &{$1.65\pm0.07$} &{$1.66\pm0.07$} &{$1.71\pm0.07$} \\
{}              &{$n$}               &{$3.30\pm0.13$} &{$3.37\pm0.12$} &{$3.36\pm0.12$} &{$3.55\pm0.12$} &{$3.26\pm0.10$} &{$3.44\pm0.11$} &{$3.68\pm0.12$} &{$3.82\pm0.13$} \\[0.5ex]
{$1 < z < 2$}   &{$R_{\rm e}$ (kpc)} &{$1.39\pm0.07$} &{$1.46\pm0.08$} &{$1.45\pm0.07$} &{$1.37\pm0.07$} &{$1.23\pm0.06$} &{$1.22\pm0.06$} &{$1.17\pm0.06$} &{$1.16\pm0.05$} \\
{}              &{$n$}               &{$2.96\pm0.14$} &{$3.41\pm0.15$} &{$3.64\pm0.13$} &{$3.70\pm0.12$} &{$3.48\pm0.13$} &{$3.51\pm0.13$} &{$3.62\pm0.13$} &{$3.66\pm0.13$} \\[0.5ex]
{$2 < z < 3$}   &{$R_{\rm e}$ (kpc)} &{$0.91\pm0.12$} &{$1.26\pm0.14$} &{$1.50\pm0.17$} &{$1.41\pm0.16$} &{$1.23\pm0.15$} &{$1.13\pm0.13$} &{$1.10\pm0.11$} &{$1.03\pm0.10$} \\
{}              &{$n$}               &{$2.19\pm0.37$} &{$2.06\pm0.30$} &{$2.21\pm0.27$} &{$2.49\pm0.28$} &{$2.34\pm0.19$} &{$2.54\pm0.21$} &{$2.36\pm0.22$} &{$2.46\pm0.23$} \\[1ex]
\multicolumn{10}{c}{\rule[0.5ex]{0.3\linewidth}{0.5pt}\quad{PSB}\quad\rule[0.5ex]{0.3\linewidth}{0.5pt}} \\[1ex]
{$0.5 < z < 1$} &{$R_{\rm e}$ (kpc)} &{$1.44\pm0.16$} &{$1.61\pm0.17$} &{$1.56\pm0.13$} &{$1.46\pm0.13$} &{$1.54\pm0.13$} &{$1.48\pm0.14$} &{$1.40\pm0.13$} &{$1.44\pm0.13$} \\
{}              &{$n$}               &{$1.99\pm0.16$} &{$1.97\pm0.21$} &{$1.97\pm0.16$} &{$2.08\pm0.22$} &{$2.19\pm0.19$} &{$2.18\pm0.22$} &{$2.13\pm0.21$} &{$2.18\pm0.21$} \\[0.5ex]
{$1 < z < 2$}   &{$R_{\rm e}$ (kpc)} &{$0.86\pm0.09$} &{$0.94\pm0.08$} &{$1.00\pm0.08$} &{$0.94\pm0.07$} &{$0.92\pm0.08$} &{$0.89\pm0.08$} &{$0.81\pm0.07$} &{$0.87\pm0.07$} \\
{}              &{$n$}               &{$3.07\pm0.23$} &{$3.41\pm0.19$} &{$3.80\pm0.20$} &{$3.68\pm0.18$} &{$3.36\pm0.18$} &{$3.42\pm0.21$} &{$3.40\pm0.19$} &{$3.33\pm0.17$} \\[0.5ex]
{$2 < z < 3$}   &{$R_{\rm e}$ (kpc)} &{$0.70\pm0.10$} &{$0.84\pm0.12$} &{$1.05\pm0.17$} &{$0.90\pm0.14$} &{$0.96\pm0.13$} &{$0.79\pm0.11$} &{$0.81\pm0.11$} &{$0.75\pm0.10$} \\
{}              &{$n$}               &{$2.52\pm0.43$} &{$2.78\pm0.28$} &{$3.17\pm0.25$} &{$3.26\pm0.26$} &{$3.69\pm0.32$} &{$3.78\pm0.33$} &{$3.63\pm0.31$} &{$3.66\pm0.29$} \\[0.5ex]
\hline
\end{tabular}
\end{minipage}
\end{table*}

\begin{figure*}
\includegraphics[width=0.85\textwidth]{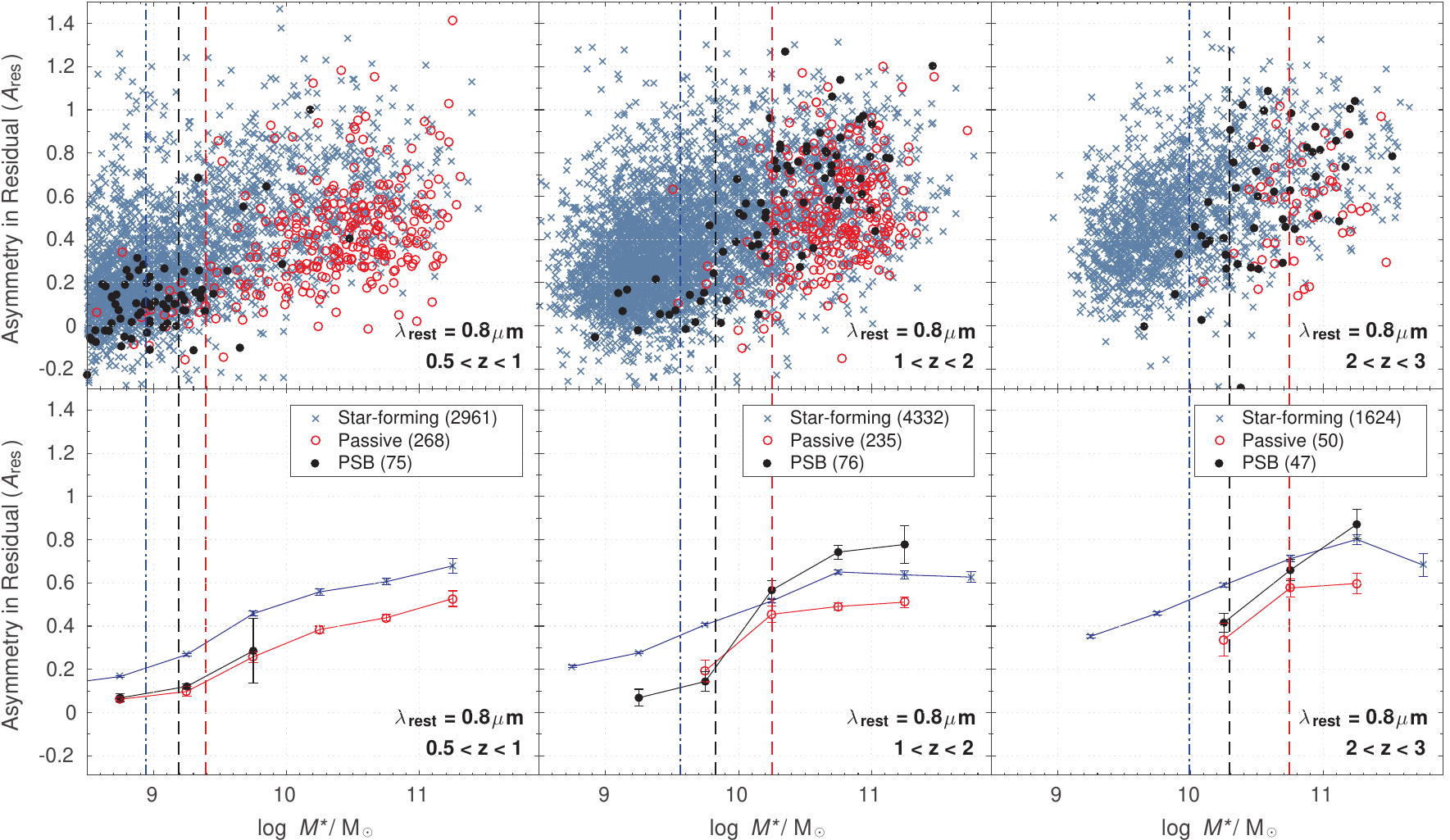}
\centering
\vspace{0.0cm}
\caption{\label{ares-vs-mass-restframe} The asymmetry in the residual ($A_{\rm res}$), determined using the PRIMER waveband closest to a rest-frame wavelength of $0.8\,\rm\mu{m}$. The layout, symbols, and binning are identical to Fig.~\ref{ares-vs-mass}, which uses observed-frame $F200W$ measurements. The resulting relations are fully consistent with those derived only from the $F200W$ imaging, indicating that the observed trends in $A_{\rm res}$ are not driven by wavelength-dependent effects.}
\end{figure*}

\bsp

\label{lastpage}

\end{document}